\documentclass[journal,letterpaper]{IEEEtran}
\usepackage{amsmath,amsfonts}
\usepackage{algorithm}
\usepackage{algpseudocode}
\usepackage{array}
\usepackage[caption=false,font=normalsize,labelfont=sf,textfont=sf]{subfig}
\usepackage{textcomp}
\usepackage{stfloats}
\usepackage{url}
\usepackage{verbatim}
\usepackage{graphicx}
\usepackage{cite}
\usepackage{float}
\usepackage{amssymb} 
\usepackage{xcolor}
\begin{document}

\title{Joint Beamforming and Position Optimization for Movable-Antenna and Movable-Element RIS–Aided Full-Duplex 6G MISO Systems}

\author{Ayda Nodel Hokmabadi and Chadi Assi \textit{~}}




\maketitle
\begin{abstract}

Full-duplex communication substantially enhances spectral efficiency by enabling simultaneous transmission and reception on the same time–frequency resources. However, its practical deployment remains hindered by strong residual self-interference and inter-user interference, which severely degrade system performance. This work investigates a full-duplex MISO network that leverages movable-antenna base stations (MA-BS) and movable-element reconfigurable intelligent surfaces (ME-RIS) to overcome these limitations in next-generation 6G systems. Unlike conventional fixed-geometry architectures, the proposed framework jointly optimizes antenna and RIS element positions, together with RIS phase shifts, to strengthen desired links while suppressing interference. Our design objective is to maximize the system sum rate through the joint optimization of transmit and receive beamforming vectors, uplink transmit powers, RIS phase shifts, and the spatial locations of both the BS antennas and RIS elements. To solve this challenging nonconvex problem, an alternating optimization algorithm is developed, employing semidefinite relaxation for beamforming design and successive convex approximation for position optimization. Simulation results demonstrate that the proposed ME-RIS–assisted architecture with movable BS antennas offers substantial gains over conventional fixed-position full-duplex networks. These findings highlight the potential of integrating movable antennas with movable RIS elements as a key enabler for high-performance full-duplex operation in future 6G wireless systems.
\end{abstract}

\begin{IEEEkeywords}
Movable antennas, movable-element reconfigurable intelligent surface (ME-RIS), Full-duplex (FD), self-interference cancellation, alternating optimization (AO).
\end{IEEEkeywords}

\section{Introduction}

Wireless communication has advanced steadily over the past few decades to accommodate ever‑increasing data rates and massive device connectivity. In emerging fifth‑ and sixth‑generation (5G/6G) networks, however, conventional solutions based on fixed antenna arrays, static beamforming, and rigid infrastructure deployments are reaching fundamental limits in terms of spectral efficiency, coverage, and robustness, particularly in dense, high‑mobility, and blockage‑prone environments. These limitations motivate architectures that can actively reshape the propagation environment, rather than merely adapting signaling strategies at fixed transmitter and receiver locations. In this context, movable antenna systems (MASs) and reconfigurable intelligent surfaces (RIS) have attracted growing interest due to their ability to dynamically tailor the wireless channel ~\cite{zhu2024movable_survey, hassouna2023ris_survey}. By physically or virtually adjusting the positions of antenna elements, movable antenna technologies can better exploit spatial diversity, alleviate blockages, and enhance link reliability in scenarios where conventional fixed arrays are inherently constrained. Similarly, RIS can manipulate the phase and, in some implementations, the amplitude response of nearly passive reflecting elements to intelligently steer and combine signals, creating programmable propagation paths that complement or bypass conventional line‑of‑sight links. Consequently, movable antenna technology introduces a new paradigm in which the network not only optimizes its transmission strategies but also actively controls the geometry and characteristics of the radio environment to overcome intrinsic limitations of current 5G systems.

Unlike traditional MIMO architectures with stationary antenna elements, movable antenna systems allow continuous repositioning of antennas within a prescribed region to exploit fine‑grained spatial variations of the wireless channel \cite{zhu2024modeling, ma2024mimo_capacity}. This additional spatial degree of freedom can yield substantial gains in spatial diversity, beamforming, and interference management. More importantly, by adjusting antenna positions, the system can deliberately seek more favorable channel realizations, rather than being confined to the performance dictated by a fixed geometry. Recent studies have demonstrated the benefits of movable antennas in a broad range of applications, including multiuser communications~\cite{zhu2024multiuser, khisa2025meta}, secure transmission~\cite{cheng2023secure_ma}, integrated sensing and communication (ISAC)~\cite{amhaz2025meta}, and full‑duplex operation~\cite{ding2025ma_fd, khisa2025meta, amhaz2025meta}. In parallel, the field‑response channel model has emerged as a standard modeling framework for such systems, capturing the dependence of channel phases on antenna positions while assuming approximately constant angles and path gains within the movement region.

Reconfigurable intelligent surfaces (RIS) constitute another key enabler for 5G/6G networks. A RIS is typically realized as a planar array of nearly passive reflecting elements whose individual phase responses can be independently controlled~\cite{renzo2021smart_radio, sharma2021ris_comprehensive}. By coherently combining reflected signals, RIS can enhance the received signal power, suppress interference, and extend coverage, all with low hardware complexity and high energy efficiency. Owing to their passive nature and the absence of conventional RF chains, RIS‑aided architectures offer a promising pathway toward energy‑sustainable networks~\cite{wu2021intelligent, liu2021reconfigurable}. Building on this concept, movable‑element reconfigurable intelligent surfaces (ME‑RIS) have been proposed as an enhanced RIS architecture in which each reflecting element can change its physical position within a designated region, in addition to its phase response~\cite{MERIS6}. By jointly optimizing the spatial positions and phase shifts of the reflecting elements, ME‑RIS introduces extra degrees of freedom for shaping the wireless propagation environment, enabling more flexible reflected paths, improved channel gains, and more effective interference control. The performance of ME‑RIS has been examined for various objectives, such as downlink sum‑rate maximization in multiuser MISO networks~\cite{MERIS2} and single‑user SISO settings~\cite{MERIS3}, typically assuming fixed‑position antennas at the base station (FPA‑BS). Furthermore, secure communication for ME‑RIS‑aided systems with FPA‑BS has been investigated by maximizing the sum secrecy rate~\cite{MERIS1}.

Indeed when deployed in isolation, both movable antennas and conventional RIS exhibit inherent limitations. Fixed‑position RIS, although phase‑reconfigurable, cannot fully adapt to spatially varying channel conditions because the physical locations of its elements remain static. Likewise, movable antennas at the base station or user equipment, while capable of mitigating severe blockages or deep fading, cannot fully exploit the additional degrees of freedom available through programmable reflections. These observations motivate the development of ME‑RIS architectures in which each reflecting element is both position‑adjustable and phase‑reconfigurable~\cite{zhao2025me_ris, zhang2024ma_ris_geometry}. This dual reconfigurability in space and phase greatly enhances the flexibility of channel shaping, allowing ME‑RIS to jointly optimize element positions and phase shifts and to synthesize cascaded channel conditions that are unattainable with conventional fixed‑position RIS~\cite{wei2024movable_ris, li2024mis}. The joint deployment of MA‑BS and ME‑RIS has therefore attracted interest, and recent works have analyzed downlink sum‑rate maximization in multiuser systems~\cite{MERIS4} and in single‑user scenarios~\cite{MERIS5}.

Full‑duplex (FD) communication offers another powerful mechanism to improve spectral efficiency by enabling simultaneous transmission and reception over the same frequency band. In practice, however, FD systems are fundamentally challenged by strong self‑interference (SI), whereby the transmitted signal leaks into the co‑located receiver and can dominate the desired signal~\cite{sabharwal2014inband, zhang2016fullduplex, khisa2025meta}. Despite substantial progress in SI suppression at the antenna, analog, and digital domains~\cite{kolodziej2019inband_survey, amjad2021antenna_sic, masmoudi2021digital_sic}, residual SI remains a primary bottleneck to realizing the theoretical gains of full‑duplex operation~\cite{kim2015survey_fd}. Moreover, FD architectures must also address inter‑user interference between uplink and downlink terminals, further increasing system design complexity~\cite{khisa2022fd_crsma, amhaz2023fd_uav_rsma}.

The joint use of movable antennas at the base station and ME‑RIS in full‑duplex systems opens up a rich design space for tackling multiple challenges simultaneously, including enhancing desired signal paths, suppressing residual self‑interference at the base station, and mitigating inter‑user interference. By jointly optimizing base‑station antenna positions, RIS element positions, and transmit/receive beamforming vectors, such systems can exploit both spatial and phase reconfiguration to a far greater extent than either technology alone. Recent studies on movable‑antenna‑assisted full‑duplex systems with rate‑splitting multiple access~\cite{khisa2025ma_fd_rsma} and integrated sensing and communication~\cite{amhaz2025isac_fd_ma} further underscore the potential of combining movable antennas with ME‑RIS as a promising direction for next‑generation 6G networks.

\subsection{Contributions}
This work investigates a full‑duplex MISO network in which a movable‑antenna base station (MA‑BS), equipped with multiple transmit and receive antennas, simultaneously serves a downlink user and an uplink user with the aid of a movable‑element reconfigurable intelligent surface (ME‑RIS).
\begin{itemize}
\item A novel system architecture is introduced, which, to the best of our knowledge, is the first to jointly exploit an MA‑BS and an ME‑RIS for full‑duplex communications, thereby enabling simultaneous spatial and phase reconfiguration of the propagation environment.
\item  A sum‑rate maximization problem is formulated that jointly optimizes the BS transmit and receive beamforming vectors, uplink transmit power, RIS phase shifts, and the three‑dimensional positions of both BS antennas and RIS elements, under quality‑of‑service constraints, transmit power budgets, and geometric feasibility constraints on antenna and element locations.
\item  An alternating optimization framework is developed to tackle the resulting highly coupled nonconvex problem, where semidefinite relaxation (SDR) combined with sequential rank‑one constraint relaxation (SROCR) is employed for beamforming design, and successive convex approximation (SCA) is used for optimizing the positions of the BS antennas and RIS elements.
\item  Numerical results demonstrate that the proposed ME‑RIS‑aided full‑duplex system with movable BS antennas achieves substantial performance gains compared with conventional full‑duplex networks employing fixed‑position antennas and RIS elements.
\end{itemize} 

\subsection{Organization}
The remainder of the manuscript is structured as follows. Section~II presents the system model, including the network architecture, signal model, and field-response-based channel model. In Section~III, the optimization problem formulation is discussed. In Section~IV, the proposed alternating optimization algorithms are explained. Finally, the numerical results and conclusion are presented in Section~V and Section~VI, respectively. 

\textit{Notation:} In this paper, the boldface lowercase and uppercase letters denote vectors and matrices, respectively. Also, $(\cdot)^T$ and $(\cdot)^H$ represent transpose and conjugate transpose. $\|\cdot\|$ denotes the Euclidean norm. $\mathbb{C}$ represents the set of complex numbers. $\text{diag}(\cdot)$ is a diagonal matrix from a vector. $\mathcal{CN}(\mu, \sigma^2)$ denotes the circularly symmetric complex Gaussian distribution with mean $\mu$ and variance $\sigma^2$. $\mathbb{E}\{\cdot\}$ denotes expectation.

\section{System Model}

\subsection{Network Model}
\begin{figure} [!t]
    \centering
    \includegraphics[width=1.0\linewidth]{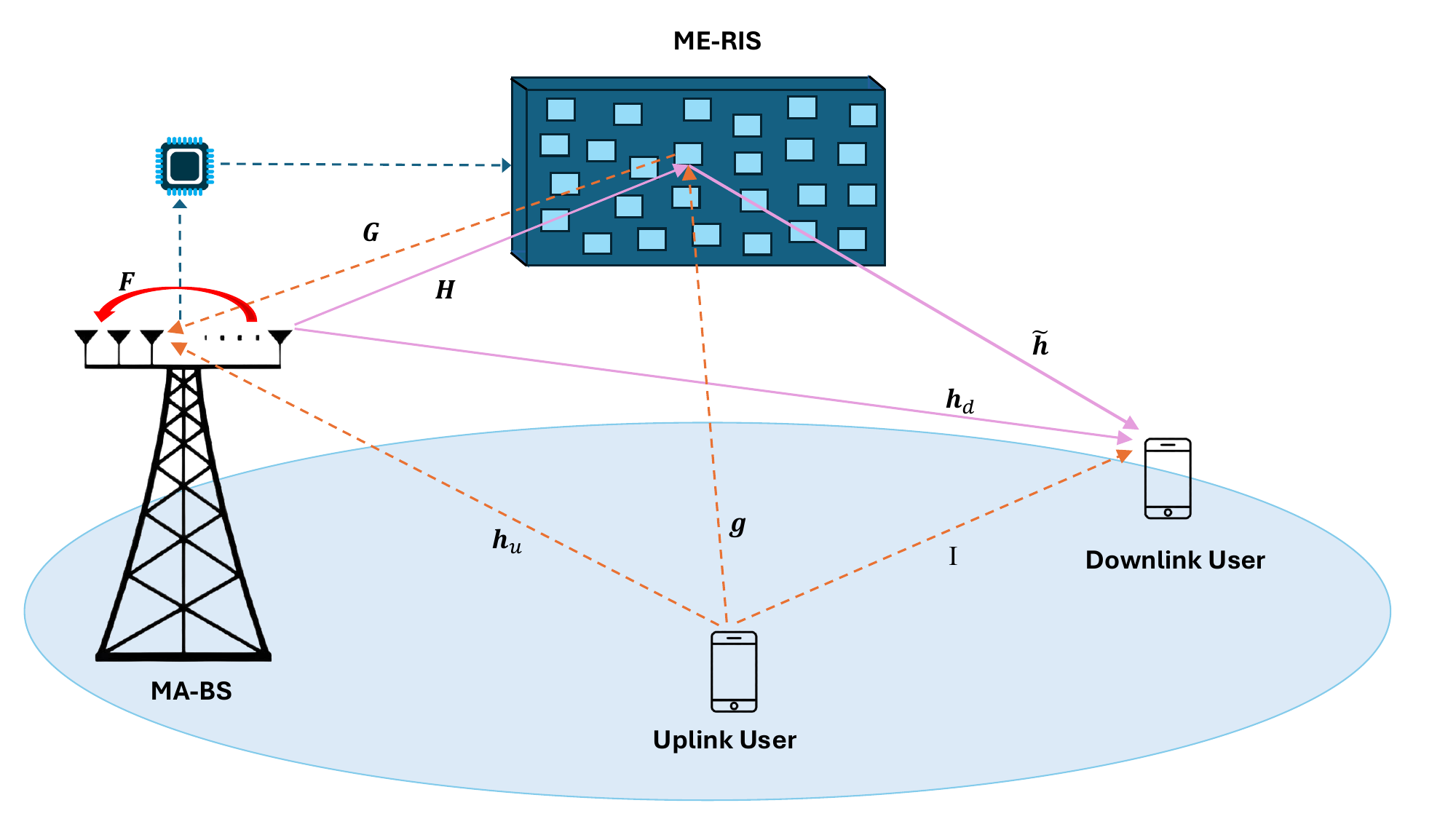}
    \caption{Full-duplex system model}
    \label{fig:system}
\end{figure}

We consider a full-duplex (FD) MISO network which consists of a base station (MA--BS) equipped with $M_t$ movable transmit antennas and $M_r$ movable receive antennas, an ME--RIS with $N$ movable passive reflecting elements, one single-antenna downlink (DL) user, and one single-antenna uplink (UL) user as shown in Fig.\ref{fig:system}. Given this system, our aim is to maximize the network sum rate by jointly optimizing the BS downlink beamforming and uplink combiner vector, UL transmit power, ME--RIS phase shifts, and antenna/elements locations at the BS and ME-RIS.\\
We denote the BS--RIS channel between the RIS and BS transmit and receive arrays as $\mathbf{H} \in \mathbb{C}^{N \times M_t}$ and $\mathbf{G} \in \mathbb{C}^{M_r \times N}$, respectively. Also, the direct BS-DL and UL-BS links are given as $\mathbf{h}_d \in \mathbb{C}^{1 \times M_t}$ and $\mathbf{h}_u \in \mathbb{C}^{M_r \times 1}$, respectively. Moreover, the RIS--DL and UL--RIS links are presented by $\tilde{\mathbf{h}} \in \mathbb{C}^{1 \times N}$ and $\mathbf{g} \in \mathbb{C}^{N \times 1}$, respectively, and the inter-user interference channel is shown by $I \in \mathbb{C}$. Finally, $\mathbf{F} \in \mathbb{C}^{M_r \times M_t}$ refers to the residual self-interference (SI) channel between the BS transmit and receive antennas. Additionally, the noise at the DL user and the BS are modeled as the circularly symmetric complex Gaussian random variables denoted by $n_d \sim \mathcal{CN}(0, \sigma_d^2)$ and $\mathbf{n}_{\mathrm{BS}} \sim \mathcal{CN}(\mathbf{0}, \sigma_u^2 \mathbf{I}_{M_r})$, respectively. Also, the parameter $\eta \in [0, 1]$ is used to represent the residual SI, where $\eta = 0$ corresponds to ideal cancellation and a larger $\eta$ shows stronger residual SI.
\subsection{Signal Model}
Let $s_d$ and $s_u$ denote the DL and UL information symbols, respectively, with $\mathbb{E}\{|s_d|^2\} = 1$ and $\mathbb{E}\{|s_u|^2\} = 1$. The BS applies the transmit beamforming vector $\boldsymbol{\omega} \in \mathbb{C}^{M_t \times 1}$ for DL transmission, while the UL user transmits its signal with power $p \in [0, P_{u}^{\text{max}}]$. In addition, the BS uses a linear receive combiner $\mathbf{v} \in \mathbb{C}^{M_r \times 1}$ to detect the UL signal. Moreover, the ME--RIS phase shift profile is defined as  a diagonal matrix of 
$\boldsymbol{\Phi} = \text{diag}(\theta_1, \ldots, \theta_N), $where $ |\theta_n| = 1, \forall n,$
and $\theta_n = e^{j\vartheta_n}$ with $\vartheta_n \in [0, 2\pi]$.

\subsubsection{Downlink Transmission}
In DL, the signal is transmitted by the BS, and the received signal at the DL user is given by
\begin{equation}
y_d = \underbrace{(\mathbf{h}_d + \tilde{\mathbf{h}} \boldsymbol{\Phi} \mathbf{H}) \boldsymbol{\omega} s_d}_{\text{desired DL signal}} + \underbrace{\sqrt{p} (I + \tilde{\boldsymbol{h}} \boldsymbol{\Phi} \boldsymbol{g} )s_u}_{\text{UL to DL interference}} + n_d.
\end{equation}
Accordingly, the downlink SINR is shown as
\begin{equation}
\gamma_{\text{DL}} = \frac{\left|(\mathbf{h}_d + \tilde{\mathbf{h}} \boldsymbol{\Phi} \mathbf{H}) \boldsymbol{\omega}\right|^2}{p|I + \tilde{\boldsymbol{h}} \boldsymbol{\Phi} \boldsymbol{g}|^2 + \sigma_d^2}.
\label{eq:sinr_dl}
\end{equation}

\subsubsection{Uplink Transmission}
In UL, the signal is transmitted by UL user, received at the BS, and is given by
\begin{equation}
\mathbf{y}_{\mathrm{BS}} = \underbrace{\sqrt{p} \left(\mathbf{h}_u + \mathbf{G}\boldsymbol{\Phi}\mathbf{g}\right) s_u}_{\text{desired UL signal}} + \underbrace{\sqrt{\eta}\left(\mathbf{F} + \mathbf{G}\boldsymbol{\Phi}\mathbf{H}\right) \boldsymbol{\omega} s_d}_{\text{SI}} + \mathbf{n}_{\mathrm{BS}}. \label{eq:sinr_ul}
\end{equation}
After linear combining, the detected UL signal is $\hat{s}_u = \mathbf{v}^H \mathbf{y}_{\mathrm{BS}}$. Thus, the uplink SINR is
\begin{equation}
\gamma_{\text{UL}} = \frac{p \left|\mathbf{v}^H \left(\mathbf{h}_u + \mathbf{G}\boldsymbol{\Phi}\mathbf{g}\right)\right|^2}{\eta \left|\mathbf{v}^H \left(\mathbf{F} + \mathbf{G}\boldsymbol{\Phi}\mathbf{H}\right) \boldsymbol{\omega}\right|^2 + \sigma_u^2 \|\mathbf{v}\|^2}.
\end{equation}

\subsubsection{Sum-rate}
The network sum rate is formulated as follows
\begin{equation}
R_{\text{sum}} = \log_2(1 + \gamma_{\text{DL}}) + \log_2(1 + \gamma_{\text{UL}}).\label{eq:sumrate}
\end{equation}

\subsection{Field-Response Based Channel Model}

All communication channels are modeled based on the field-response channel model~\cite{zhu2023field_response, ma2024mimo_capacity}. More specifically, in movable-antenna (MA) systems, the channel response is affected by both the propagation environment and the position of the antennas. Following~\cite{zhu2024modeling}, it is assumed that the movable region for each antenna/element is sufficiently small compared to the transmitter-receiver distance, which ensures far-field propagation at both ends. This assumption is reasonable in practice since MA-BS and ME-RIS moving ranges are typically limited to few wavelengths. Therefore, a plane-wave approximation is applicable. The angles of departure/arrival (AoDs/AoAs) and the path-gain magnitudes remain approximately invariant over the movement region, while the phases of multipath components vary with the antenna/element positions. This implies the key benefit of MA/ME-RIS architectures, namely, the ability to reshape the channel phases by position adaptation. Now, for any link in the system with $L$ dominant propagation paths, the channel matrix can be formulated using the field response model as
\begin{equation}
\mathbf{C} = \mathbf{F}^H \boldsymbol{\Sigma} \mathbf{E},
\end{equation}
where $\mathbf{E} \in \mathbb{C}^{L \times M}$ is the transmit field-response matrix (FRM), $\mathbf{F} \in \mathbb{C}^{L \times N}$ is the receive FRM, and $\boldsymbol{\Sigma} \in \mathbb{C}^{L \times L}$ is the path-response matrix (PRM) that shows the complex path gains.
The path-response matrix $\boldsymbol{\Sigma}$ is a diagonal matrix
\begin{equation}
\boldsymbol{\Sigma} = \text{diag}(\zeta_1, \zeta_2, \ldots, \zeta_L),
\end{equation}
where $\zeta_\ell$ represents the complex path gain of the $\ell$-th propagation path. For all links in our system, we model the path gains as independent circularly symmetric complex Gaussian (CSCG) random variables:
\begin{equation}
\zeta_\ell \sim \mathcal{CN}\left(0, \frac{\beta_0 d^{-\alpha}}{L}\right), \quad \ell = 1, 2, \ldots, L,
\end{equation}
where $\beta_0$  denotes the path loss at the reference distance of
$1m$, $\alpha$ is the path loss exponent, and $d$ represents the distance between the transmitter and the receiver. 

\subsubsection{Downlink channels}
In the downlink, let $L_{\text{BR}}$, $L_{\text{Bd}}$, and $L_{\text{Rd}}$ denote the numbers of paths for the BS--RIS, BS--DL, and RIS--DL links, respectively. We consider local coordinate systems $x_\text{B}$-$O_\text{B}$-$y_\text{B}$ at the BS and $x_\text{S}$-$O_\text{S}$-$y_\text{S}$ at the RIS with reference points $O_\text{B} = [0, 0]^T$ and $O_\text{S} = [0, 0]^T$.

Let $\mathbf{t}_m = [x_m, y_m]^T$ denote the position of the $m$-th BS transmit antenna and $\mathbf{r}_n = [x_n, y_n]^T$ the position of the $n$-th RIS element. We define the transmit-antenna position matrix (APM) and RIS element position matrix (EPM) as
\begin{equation}
\mathbf{T}_t = [\mathbf{t}_1, \ldots, \mathbf{t}_{M_t}], \quad \mathbf{R} = [\mathbf{r}_1, \ldots, \mathbf{r}_N].
\end{equation}

 Let $\theta_\ell^{\text{BR, B}}$ and $\phi_\ell^{\text{BR, B}}$ denote the elevation and azimuth AoDs of the $\ell$-th BS--RIS path, respectively. The transmit field-response vector (FRV) of the $m$-th BS antenna is
\begin{equation}
\mathbf{e}_{\text{BR}}(\mathbf{t}_m) = \left[e^{j \frac{2\pi}{\lambda} \rho_1^{\text{BR, B}}(\mathbf{t}_m)}, \ldots, e^{j \frac{2\pi}{\lambda} \rho_{L_{\text{BR}}}^{\text{BR, B}}(\mathbf{t}_m)}\right]^T,
\end{equation}
where the path-length difference relative to the reference point is
$\rho_\ell^{\text{BR, B}}(\mathbf{t}_m) = x_m \cos(\theta_\ell^{\text{BR, B}}) \sin(\phi_\ell^{\text{BR, B}}) + y_m \sin(\theta_\ell^{\text{BR, B}}).$
Stacking all FRVs yields the BS transmit FRM
\begin{equation}
\mathbf{E}_{\text{BR}} = \left[\mathbf{e}_{\text{BR}}(\mathbf{t}_1), \ldots, \mathbf{e}_{\text{BR}}(\mathbf{t}_{M_t})\right] \in \mathbb{C}^{L_{\text{BR}} \times M_t}.
\end{equation}

Similarly, let $\theta_\ell^{\text{BR, S}}$ and $\phi_\ell^{\text{BR, S}}$ denote the elevation and azimuth AoAs of the $\ell$-th BS--RIS path at the RIS. The incident FRV at the $n$-th RIS element is
\begin{equation}
\mathbf{f}_{\text{BR}}(\mathbf{r}_n) = \left[e^{j \frac{2\pi}{\lambda} \rho_1^{\text{BR, S}}(\mathbf{r}_n)}, \ldots, e^{j \frac{2\pi}{\lambda} \rho_{L_{\text{BR}}}^{\text{BR, S}}(\mathbf{r}_n)}\right]^T,
\end{equation}
with
$\rho_\ell^{\text{BR, S}}(\mathbf{r}_n) = x_n \cos(\theta_\ell^{\text{BR, S}}) \sin(\phi_\ell^{\text{BR, S}}) + y_n \sin(\theta_\ell^{\text{BR, S}}).$

The RIS incident FRM is then
\begin{equation}
\mathbf{F}_{\text{BR}} = \left[\mathbf{f}_{\text{BR}}(\mathbf{r}_1), \ldots, \mathbf{f}_{\text{BR}}(\mathbf{r}_N)\right] \in \mathbb{C}^{L_{\text{BR}} \times N}.
\end{equation}

Let $\boldsymbol{\Sigma}_{\text{BR}} \in \mathbb{C}^{L_{\text{BR}} \times L_{\text{BR}}}$ denote the PRM with diagonal elements. The BS--RIS channel is given as
\begin{equation}
\mathbf{H} = \mathbf{F}_{\text{BR}}^H \boldsymbol{\Sigma}_{\text{BR}} \mathbf{E}_{\text{BR}} \in \mathbb{C}^{N \times M_t}.
\label{FR_construction}
\end{equation}

For the BS--DL direct channel, let $\boldsymbol{\sigma}_{\text{Bd}} \in \mathbb{C}^{1 \times L_{\text{Bd}}}$ denote the PRV from $O_\text{B}$ to the DL user with elements following $\mathcal{CN}\left(0, \frac{\beta_0 d^{-\alpha_\text{Bd}}}{L_{\text{Bd}}}\right)$. Then
\begin{equation}
\mathbf{h}_d = \boldsymbol{\sigma}_{\text{Bd}} \mathbf{E}_{\text{Bd}},
\end{equation}
where $\mathbf{E}_{\text{Bd}} \in \mathbb{C}^{L_{\text{Bd}} \times M_t}$ is obtained by constructing the corresponding FRVs using the AoDs of the BS--DL paths.

 Similarly, the RIS--DL channel is modeled as
$\tilde{\mathbf{h}} = \boldsymbol{\sigma}_{\text{Rd}} \mathbf{E}_{\text{Rd}},$
where $\boldsymbol{\sigma}_{\text{Rd}} \in \mathbb{C}^{1 \times L_{\text{Rd}}}$ and $\mathbf{E}_{\text{Rd}} \in \mathbb{C}^{L_{\text{Rd}} \times N}$ are defined using the ME--RIS to DL path angles.

\subsubsection{Uplink channels}
The uplink channels are modeled similarly. Let $L_{\text{RB}}$, $L_{\text{Bu}}$, and $L_{\text{Ru}}$ denote the numbers of paths between the RIS--BS, UL--BS, and UL--RIS links, respectively. The BS receiver antenna positions are denoted by $\mathbf{u}_m = [x_m, y_m]^T$ and collected as
$\mathbf{T}_r = [\mathbf{u}_1, \ldots, \mathbf{u}_{M_r}].$
Using the same formulation as in Eq.~\eqref{FR_construction}, the RIS--BS channel is given as
$\mathbf{G} = \mathbf{F}_{\text{RB}}^H \boldsymbol{\Sigma}_{\text{RB}} \mathbf{E}_{\text{RB}} \in \mathbb{C}^{M_r \times N},$
where $\boldsymbol{\Sigma}_{\text{RB}} \in \mathbb{C}^{L_{\text{RB}} \times L_{\text{RB}}}$ is the PRM, and $\mathbf{E}_{\text{RB}}$, $\mathbf{F}_{\text{RB}}$ are computed from the BS receive APM, which is $\mathbf{T}_r$, and RIS EPM, which is $\mathbf{R}$.

Furthermore, the direct UL--BS channel is formulated as
\begin{equation}
 \mathbf{h}_u = \mathbf{F}_{\text{Bu}}^H \boldsymbol{\sigma}_{\text{Bu}},   
\end{equation}
where $\boldsymbol{\sigma}_{\text{Bu}} \in \mathbb{C}^{L_{\text{Bu}} \times 1}$ and $\mathbf{F}_{\text{Bu}} \in \mathbb{C}^{L_{\text{Bu}} \times M_r}$ is generated from the AoAs at the BS receive array. Additionally, the UL--RIS channel is modeled as
\begin{equation}
    \mathbf{g} = \mathbf{E}_{\text{Ru}}^H \boldsymbol{\sigma}_{\text{Ru}},
\end{equation}
where $\boldsymbol{\sigma}_{\text{Ru}} \in \mathbb{C}^{L_{\text{Ru}} \times 1}$ and $\mathbf{E}_{\text{Ru}} \in \mathbb{C}^{L_{\text{Ru}} \times N}$ are obtained using the UL--RIS path angles and the RIS EPM, $\mathbf{R}$.
Also, the residual SI channel $\mathbf{F} \in \mathbb{C}^{M_r \times M_t}$ between the BS transmit and receive arrays is modeled similarly using the FR framework in Eq.~\eqref{FR_construction}.

\section{Problem Formulation}
In order to achieve the potential of the FD communication, we aim to maximize the sum rate of the DL and UL users by jointly optimizing the transmitting and receiving beamforming vectors at the BS,  the positions of transmitting and receiving antennas, the phase shift profile matrix and element positions at ME-RIS, and the uplink transmit power subject to QoS constraints (minimum data rate) for both users, maximum transmit power constraints at the BS and UL user, unit-modulus constraint for the ME-RIS phase shifting elements, the movable range for each movable antenna at BS or element at ME-RIS, and the minimum distance constraint between each two movable antennas/elements. Therefore, we formulate the optimization problem as follows:
\begin{align}
\max_{\mathcal{X}} \quad 
& \log_2\!\left(1+\gamma_{\mathrm{DL}}\right)
+ \log_2\!\left(1+\gamma_{\mathrm{UL}}\right) 
\label{P0:obj} \\[2mm]
\text{s.t.}\quad
& \|\boldsymbol{\omega}\|_2^2 \le P_{\mathrm{BS}}^{\max}, 
\label{P0:BSpower} \\
& 0 \le p \le P_u^{\max}, 
\label{P0:ULpower} \\
& R_{\mathrm{DL}} \ge R^{th}_{\mathrm{DL}}, 
\label{P0:DLQoS} \\
& R_{\mathrm{UL}} \ge R^{th}_{\mathrm{UL}}, 
\label{P0:ULQoS} \\
& |[\boldsymbol{\Phi}]_{n,n}| = 1, \quad \forall n \in \{1,\ldots,N\}, 
\label{P0:unitmod} \\
& \mathbf{r}_n \in \mathcal{R}_n,\quad \forall n \in \{1,\ldots,N\}, 
\label{P0:Rbox} \\
& \mathbf{t}_m \in \mathcal{T}_m,\quad \forall m \in \{1,\ldots,M_t\}, 
\label{P0:Ttbox} \\
& \mathbf{u}_m \in \mathcal{U}_m,\quad \forall m \in \{1,\ldots,M_r\}, 
\label{P0:Trbox} \\
& \|\mathbf{r}_n - \mathbf{r}_{n'}\|_2 \ge d_0, \quad \forall n \neq n', 
\label{P0:Rsep} \\
& \|\mathbf{t}_m - \mathbf{t}_{m'}\|_2 \ge d_0, \quad \forall m \neq m', 
\label{P0:Ttsep} \\
& \|\mathbf{u}_m - \mathbf{u}_{m'}\|_2 \ge d_0, \quad \forall m \neq m'. 
\label{P0:Trsep}
\end{align}
where $\mathcal{X} = \{p, \boldsymbol{\omega}, \boldsymbol{v}, \boldsymbol{\Phi}, \mathbf{T}_t, \mathbf{T}_r, \mathbf{R}\}$ is the set of the optimization variables. Also, constraint~\eqref{P0:BSpower} limits the transmit power of the base station,
\eqref{P0:ULpower} bounds the uplink transmit power,
and \eqref{P0:DLQoS}–\eqref{P0:ULQoS} guarantee the minimum QoS requirements for the DL and UL users, respectively.
Constraint~\eqref{P0:unitmod} is the unit-modulus phase-shift of the passive ME-RIS.
Constraints~\eqref{P0:Rbox}–\eqref{P0:Trbox} restrict the locations of the RIS elements,
transmit antennas and receive antennas to their feasible regions.
Finally, constraints~\eqref{P0:Rsep}–\eqref{P0:Trsep} are minimum inter-element distances $d_0$
to avoid physical overlap.

\section{Proposed AO-based Algorithms}
In this section, we divide the main optimization problem into six subproblems and optimize them alternatively. More specifically, we employ the SDR-SROCR algorithm for the BS downlink beamforming vector optimization ($\boldsymbol{\omega}$) and obtain a closed-form solution for the uplink beamforming encoder ($\boldsymbol{v}$), and the uplink transmit power. Subsequently, the BS transmitting and receiving antenna positions, in addition to the ME-RIS element positions and the phase shift profile, are optimized using the first-order SCA technique. Each subproblem and the corresponding optimization algorithm are discussed below.

\subsection{Downlink beamforming vector at BS ($\boldsymbol{\omega}$)}
We optimize the downlink transmit beamforming vector at the full--duplex BS,
denoted by $\boldsymbol{\omega}\in\mathbb{C}^{M_t \times 1}$,
while keeping other optimization variables, $\{\boldsymbol{v}, p, \boldsymbol{\Phi}, \mathbf{R}, \mathbf{T}_t, \mathbf{T}_r\}$,  fixed.
Since $\boldsymbol{\omega}$ directly determines the downlink SINR and also affects the uplink SINR through residual (SI) at the BS, the beamforming design must satisfy both downlink and uplink QoS constraints.
Accordingly, the downlink beamforming subproblem is formulated as
\begin{subequations}\label{eq:sub_w}
\begin{align}
\max_{\boldsymbol{\omega}} \quad &
\log_2\!\left(1+\gamma_{\mathrm{DL}}\right)
+
\log_2\!\left(1+\gamma_{\mathrm{UL}}\right)
\label{eq:sub_w_obj}
\\
\text{s.t.}\quad
&
\frac{
\big|
\big(\boldsymbol{h}_{\mathrm{d}}
+\tilde{\boldsymbol{h}}\boldsymbol{\Phi}\boldsymbol{H}\big)
\boldsymbol{\omega}
\big|^2
}{
p\,\big|
I
+\tilde{\boldsymbol{h}}\boldsymbol{\Phi}\boldsymbol{g}
\big|^2
+\sigma_d^2
}
\ge \gamma_{\mathrm{DL}}^{\min},
\label{eq:sub_w_DL}
\\
&
\frac{
p\,\big|
\boldsymbol{v}^{H}
\big(\boldsymbol{h}_{\mathrm{u}}
+\boldsymbol{G}\boldsymbol{\Phi}\boldsymbol{g}\big)
\big|^2
}{
\eta\,
\big|
\boldsymbol{v}^{H}
\big(\boldsymbol{F}
+\boldsymbol{G}\boldsymbol{\Phi}\boldsymbol{H}\big)
\boldsymbol{\omega}
\big|^2
+\sigma_u^2\|\boldsymbol{v}\|_2^2
}
\ge \gamma_{\mathrm{UL}}^{\min},
\label{eq:sub_w_UL}
\\
&
\|\boldsymbol{\omega}\|_2^2
\le P^{\text{BS}}_{\max}.
\label{eq:sub_w_power}
\end{align}
\end{subequations}
where $\gamma_{\mathrm{DL}}^{\min} = 2^{R^{th}_{\mathrm{DL}}} - 1$ and $\gamma_{\mathrm{UL}}^{\min} = 2^{R^{th}_{\mathrm{UL}}} - 1$.
Problem~\eqref{eq:sub_w} is non--convex due to the fractional SINR expressions
and the quadratic SI term in the uplink constraint.
To make the problem tractable, we first apply the Lagrangian dual transform (LDT)\cite{FP1}.

\textbf{\textit{Lemma~1:}} According to the LDT method introduced in \cite{FP1}, the sum-logarithms term in the sum rate expression of the system represented by $ \sum_{i \in \mathcal{S}} \log_2(1 + \gamma_{i})$, for $\mathcal{S} = \{{\mathrm{DL}}, {\mathrm{UL}}\}$ in this problem, with a general form of $\gamma_{i}=\frac{A_{i}(\boldsymbol{x})}{B_{i}(\boldsymbol{x})}$, can be transformed into a new function in the form of the sum of ratios by introducing a group of new auxiliary variables, $\boldsymbol{\zeta}=[\zeta_{1}, \zeta_{2}, \dots, \zeta_{|\mathcal{S}|}]$, and the new objective function becomes,
\begin{equation}
    f(\boldsymbol{x}, \boldsymbol{\zeta})=\sum_{i \in \mathcal{S} } \left(\log_2\left(1 + \zeta_{i}\right)-\zeta_{i}\right) +\sum_{i \in \mathcal{S}}(1+\zeta_{i}) f_{i}(\boldsymbol{x})\label{LDT1}
\end{equation}
where $\boldsymbol{x}$ is the optimization variables, and $f_{i}(\boldsymbol{x}) = \frac{A_{i}(\boldsymbol{x})}{B_{i}(\boldsymbol{x})+A_{i}(\boldsymbol{x})}$. This new defined objective function, $f(\boldsymbol{x}, \zeta_{i})$, is equivalent to the original sum rate function in problem~\eqref{eq:sub_w}, $ \sum_{i \in \mathcal{S}} \log_2(1 + \gamma_{i})$, if and only if when $\zeta_{i}=\gamma_{i}$. 

\textbf{\textit{Proof:}} 
The proof is given in Appendix~\ref{appendix1}. \hfill$\blacksquare$ 

Subsequently, after applying the LDT transformation, the transformed objective function is in the form of multiple-ratio fractional programming (MRFP). Therefore, according to the \cite{FP1} and \cite{FP2}, the quadratic transform (QT) can be applied in order to recast the objective function to a more solvable one, and this transformation is stated as follows.

\textbf{\textit{Lemma~2:}}
In the QT, a group of new auxiliary variables are introduced as $\boldsymbol{\beta}=[\beta_{1}, \beta_{2}, \dots, \beta_{|\mathcal{S}|}]$, and the MRFP term of the previous part, $\sum_{i \in \mathcal{S}} f_{i}(\boldsymbol{x})$, is transformed and written as
\begin{equation}\label{QT1}
\begin{aligned}
g(\boldsymbol{x}, \boldsymbol{\zeta}, \boldsymbol{\beta})
&= \sum_{i \in \mathcal{S}}
\Big(
2\sqrt{1+\zeta_{i}}
\,\Re\!\left\{\beta^*_{i}\sqrt{A_{i}(\boldsymbol{x})}\right\} \\
&\hspace{1.5em}
- |\beta_{i}|^2
\big(B_{i}(\boldsymbol{x})+A_{i}(\boldsymbol{x})\big)
\Big).
\end{aligned}
\end{equation}

The $ g(\boldsymbol{x}, \boldsymbol{\zeta}, \boldsymbol{\beta})$ function is equivalent to the $\sum_{i \in \mathcal{S}} f_{i}(\boldsymbol{x})$ if and only if when 
\begin{equation}
    \beta_{i} = \sqrt{1+\zeta_{i}}\frac{\sqrt{A_{i}(\boldsymbol{x})}}{B_{i}(\boldsymbol{x})+A_{i}(\boldsymbol{x})}
\end{equation}.

\textbf{\textit{Proof:}} 
The proof is given in Appendix~\ref{appendix2}. \hfill$\blacksquare$

By applying the LDT and QT to problem~\eqref{eq:sub_w}, the objective reduces to the quadratic form of 
\begin{equation}\label{eq:QT1} 
2\Re\!\left\{
\boldsymbol{a}\boldsymbol{\omega}
\right\}
-
\boldsymbol{\omega}^H\boldsymbol{B}_{\mathrm{UL}}\boldsymbol{\omega}
+ c,
\end{equation}
where
\begin{align}
\boldsymbol{a}
&=
\beta_{\mathrm{DL}}^*\sqrt{1+\zeta_{\mathrm{DL}}}\,
\boldsymbol{h}_{\mathrm{DL}},
\\
\boldsymbol{h}_{\mathrm{DL}}
&=
\boldsymbol{h}_{\mathrm{d}}
+
\tilde{\boldsymbol{h}} \boldsymbol{\Phi}\boldsymbol{H},
\\
\boldsymbol{B}_{\mathrm{UL}}
&=
\eta\,|\beta_{\mathrm{UL}}|^2\,
\big(\boldsymbol{F}+\boldsymbol{G}\boldsymbol{\Phi}\boldsymbol{H}\big)^H
\boldsymbol{v}\boldsymbol{v}^H
\big(\boldsymbol{F}+\boldsymbol{G}\boldsymbol{\Phi}\boldsymbol{H}\big).
\end{align}

and $c$ is the summation of all constant terms independent of $\boldsymbol{\omega}$.

By defining the augmented beamforming vector
$\bar{\boldsymbol{\omega}}=[\boldsymbol{\omega}^T\;1]^T$
and the rank-one matrix
$\boldsymbol{W}=\bar{\boldsymbol{\omega}}\bar{\boldsymbol{\omega}}^H$,
problem~\eqref{eq:sub_w} can be reformulated as the following
rank--constrained semidefinite program
\begin{subequations}\label{eq:SDP}
\begin{align}
\max_{\boldsymbol{W}} \quad &
\mathrm{tr}\!\left(\boldsymbol{A}\boldsymbol{W}\right)
\\
\text{s.t.}\quad
&
\mathrm{tr}\!\left(\boldsymbol{E}_{\mathrm{DL}}\boldsymbol{W}\right)
\ge b_{\mathrm{DL}},
\\
&
\mathrm{tr}\!\left(\boldsymbol{E}_{\mathrm{UL}}\boldsymbol{W}\right)
\ge b_{\mathrm{UL}},
\\
&
\mathrm{tr}(\boldsymbol{W}) \le P^{\text{BS}}_{\max}+1,
\\
&
[\boldsymbol{W}]_{M+1,M+1}=1,
\\
&
\boldsymbol{W}\succeq\boldsymbol{0},
\qquad
\mathrm{rank}(\boldsymbol{W})=1,
\end{align}
\end{subequations}
where the matrices are defined as
\begin{equation}
\boldsymbol{A}=
\begin{bmatrix}
-\boldsymbol{B}_{\mathrm{UL}} & \boldsymbol{a}^H\\
\boldsymbol{a} & 0
\end{bmatrix},
\quad
\boldsymbol{h}_{\mathrm{UL}}
=
\sqrt{\eta}\,\boldsymbol{v}^{H}
\big(\boldsymbol{F}
+\boldsymbol{G}\boldsymbol{\Phi}\boldsymbol{H}\big),
\end{equation}
\begin{equation}
\boldsymbol{E}_{\mathrm{DL}}=
\begin{bmatrix}
\boldsymbol{h}_{\mathrm{DL}}^H \boldsymbol{h}_{\mathrm{DL}} & \boldsymbol{0}\\
\boldsymbol{0} & 0
\end{bmatrix},
\qquad
\boldsymbol{E}_{\mathrm{UL}}=
\begin{bmatrix}
-\gamma_{\mathrm{UL}}^{\min}\boldsymbol{h}_{\mathrm{UL}}^H \boldsymbol{h}_{\mathrm{UL}} & \boldsymbol{0}\\
\boldsymbol{0} & 0
\end{bmatrix}.
\end{equation}
and $b_{\mathrm{DL}}$ and $b_{\mathrm{UL}}$ are the summation of all constant terms independent of $\boldsymbol{\omega}$ in \eqref{eq:sub_w_DL} and \eqref{eq:sub_w_UL}, respectively.

Since problem~\eqref{eq:SDP} is non--convex due to the rank--one constraint, we first drop the rank constraint to obtain a standard SDR problem,
and then apply the sequential rank--one constraint relaxation (SROCR) method \cite{SROCR1}.
At iteration $n$, the rank--one constraint is approximated by
\begin{equation}\label{eq:srocr}
\bar{\boldsymbol{u}}(\boldsymbol{W}^{(n)})^H
\boldsymbol{W}
\bar{\boldsymbol{u}}(\boldsymbol{W}^{(n)})
\ge
m^{(n)}\mathrm{tr}(\boldsymbol{W}),
\end{equation}
where $\bar{\boldsymbol{u}}(\boldsymbol{W}^{(n)})$ is the dominant eigenvector
of $\boldsymbol{W}^{(n)}$ and $m^{(n)}\in[0,1]$ is gradually increased.
After convergence, a near rank--one solution $\boldsymbol{W}^\star$ is obtained,
from which the beamforming vector $\boldsymbol{\omega}^\star$
is recovered from the principal eigenvector of $\boldsymbol{W}^\star$. The algorithm of the SROCR method is explained in detail in Algorithm~\ref{alg:srocr}.
\begin{algorithm}[!t]
\caption{Rank reduction via the SROCR method}
\label{alg:srocr}
\begin{small}
\begin{algorithmic}[1]
\State \textbf{Initialize:} set $n=0$, choose a feasible $\boldsymbol{\omega}^{(0)}$, threshold $\epsilon_{1}$, and maximum iteration $I_{\mathrm{SROCR}}$.
\State \textbf{Relaxed solution:} solve \eqref{eq:SDP} with the relaxed rank-one constraint to obtain $\boldsymbol{W}^{(0)}$, and set $m^{(0)}=0$.
\State Choose $\delta^{(0)} \in \left(0,\,1-\frac{\lambda_{\max}(\boldsymbol{W}^{(0)})}{\mathrm{tr}\,\!\left(\boldsymbol{W}^{(0)}\right)}\right)$.

\Repeat
    \State Solve \eqref{eq:SDP} by replacing the rank-one constraint with \eqref{eq:srocr} using $(\boldsymbol{W}^{(n)}, m^{(n)})$.
    \If{the problem is feasible}
        \State Obtain $\boldsymbol{W}^{(n+1)}$ and set $\delta^{(n+1)} \leftarrow \delta^{(n)}$.
    \Else
        \State Set $\boldsymbol{W}^{(n+1)} \leftarrow \boldsymbol{W}^{(n)}$ and $\delta^{(n+1)} \leftarrow \delta^{(n)}/3$.
    \EndIf
    \State Update
    $m^{(n+1)}=\min\!\left(1,\frac{\lambda_{\max}(\boldsymbol{W}^{(n+1)})}{\mathrm{tr}\,\!\left(\boldsymbol{W}^{(n+1)}\right)}+\delta^{(n+1)}\right)$.
    \State $n \leftarrow n+1$.
\Until{$|1-m^{(n)}|\le \epsilon_{1}$ \textbf{or} $n=I_{\mathrm{SROCR}}$}

\State \textbf{Output:} $\boldsymbol{W}^{\star}\leftarrow \boldsymbol{W}^{(n)}$, recover $\boldsymbol{\omega}^{\star}$ from $\boldsymbol{W}^{\star}$.
\end{algorithmic}
\end{small}
\end{algorithm}
\subsection{Uplink combiner vector at BS ($\boldsymbol{v}$)}
We optimize the uplink receive combining vector at the BS, denoted by $\boldsymbol{v}$, while keeping $\{\boldsymbol{\omega}, p, \boldsymbol{\Phi}, \mathbf{R}, \mathbf{T}_t, \mathbf{T}_r\}$ fixed. For the considered single--user uplink, the objective is to maximize the uplink SINR in the presence of residual self--interference and receiver noise. This leads to a generalized Rayleigh quotient structure \cite{rayleigh1}.

The corresponding subproblem is
\begin{equation}\label{eq:sub_v_original}
\begin{aligned}
\max_{\boldsymbol{v}} \quad &
\frac{p\,|\boldsymbol{v}^H (\boldsymbol{h}_u + \boldsymbol{G} \boldsymbol{\Phi} \boldsymbol{g})|^2}
{\eta|\boldsymbol{v}^H (\boldsymbol{F}  + \boldsymbol{G}  \boldsymbol{\Phi} \boldsymbol{H}) \boldsymbol{\omega} |^2 + \sigma_u^2 \|\boldsymbol{v} \|^2} \\
\text{s.t.}\quad &  
\frac{p\,|\boldsymbol{v}^H (\boldsymbol{h}_u + \boldsymbol{G} \boldsymbol{\Phi} \boldsymbol{g})|^2}
{\eta|\boldsymbol{v}^H (\boldsymbol{F}  + \boldsymbol{G}  \boldsymbol{\Phi} \boldsymbol{H}) \boldsymbol{\omega} |^2 + \sigma_u^2 \|\boldsymbol{v} \|^2} \geq \gamma^{\mathrm{UL}}_{\mathrm{min}}.
\end{aligned}
\end{equation}

The effective UL channel and the residual SI vector are:
\begin{align}
\boldsymbol{a} &\triangleq \boldsymbol{h}_u + \boldsymbol{G}\boldsymbol{\Phi}\boldsymbol{g}, \\
\boldsymbol{b} &\triangleq (\boldsymbol{F} + \boldsymbol{G}\boldsymbol{\Phi}\boldsymbol{H})\boldsymbol{\omega}.
\end{align}
Then, letting
$\boldsymbol{A} \triangleq \boldsymbol{a}\boldsymbol{a}^H, 
\boldsymbol{S} \triangleq \eta\,\boldsymbol{b}\boldsymbol{b}^H,$
the UL SINR can be written as the generalized Rayleigh quotient
\begin{equation}
\gamma_{\mathrm{UL}}=
\frac{p\,\boldsymbol{v}^H \boldsymbol{A}\boldsymbol{v}}
{\boldsymbol{v}^H(\boldsymbol{S}+\sigma_u^2\boldsymbol{I}_M)\boldsymbol{v}}.
\end{equation}
Therefore, the optimal combiner is given by the dominant generalized eigenvector \cite{rayleigh2, rayleigh3}:
\begin{equation}\label{eq:v_star}
\boldsymbol{v}^{\star} = 
\frac{\!(\sigma_u^2 \boldsymbol{I}_M + \boldsymbol{S})^{-1}\boldsymbol{a}}
{\left\|\!(\sigma_u^2 \boldsymbol{I}_M + \boldsymbol{S})^{-1}\boldsymbol{a}\right\|_2}.
\end{equation}
Finally, we verify whether the QoS constraint in \eqref{eq:sub_v_original} is satisfied using $\boldsymbol{v}^{\star}$. If the QoS constraint is violated after updating $\boldsymbol{v}$, then the current values of $\{\boldsymbol{\omega}, p, \boldsymbol{\Phi}, \mathbf{R}, \mathbf{T}_t, \mathbf{T}_r\}$ do not yield a feasible uplink QoS, and the AO continues by updating the remaining variables.

\subsection{Uplink transmit power ($p$)}
We optimize the UL transmit power $p$ while fixing $\{\boldsymbol{v}, \boldsymbol{\omega}, \boldsymbol{\Phi}, \mathbf{R}, \mathbf{T}_t, \mathbf{T}_r\}$. The variable $p$ exists in both the DL and the UL SINRs, which creates a fundamental trade--off in the sum--rate objective. Consequently, the power subproblem is a one--dimensional nonconvex optimization, but it can be solved efficiently since it involves only a single variable.

The power subproblem is formulated as
\begin{equation}\label{eq:sub_p_original}
\begin{aligned}
\max_{0\le p\le P^{\mathrm{UL}}_{\max}} \quad &
\log_2\!\left(1+\gamma_{\mathrm{DL}}\right)
+ \log_2\!\left(1+\gamma_{\mathrm{UL}}\right).
\end{aligned}
\end{equation}

The feasible set can be further restricted by the QoS constraints $\gamma_{\mathrm{DL}}\ge \gamma^{\mathrm{DL}}_{\mathrm{min}}$ and $\gamma_{\mathrm{UL}}\ge \gamma^{\mathrm{UL}}_{\mathrm{min}}$ evaluated at the current $(\boldsymbol{\omega},\boldsymbol{v},\boldsymbol{\Phi})$. Since \eqref{eq:sub_p_original} is one--dimensional, we solve it using a bisection search over the feasible interval.

\subsection{ME--RIS phase shift profile ($\boldsymbol{\Phi}$)}
We optimize the ME--RIS phase shift profile $\boldsymbol{\Phi}$ for fixed $\{\boldsymbol{v}, p, \boldsymbol{\omega}, \mathbf{R}, \mathbf{T}_t, \mathbf{T}_r\}$. The phase design is constrained by unit--modulus elements, which makes the problem highly non--convex. The phase optimization subproblem is
\begin{equation}\label{eq:sub_phi_original}
\begin{aligned}
\max_{\boldsymbol{\Phi}} \quad &
\log_2\!\left(1+\gamma_{\mathrm{DL}}\right)
+ \log_2\!\left(1+\gamma_{\mathrm{UL}}\right) \\
\text{s.t.}\quad &
\gamma_{\mathrm{DL}}(\boldsymbol{\Phi}) \ge \gamma^{\mathrm{DL}}_{\mathrm{min}}, \\
&
\gamma_{\mathrm{UL}}(\boldsymbol{\Phi}) \ge \gamma^{\mathrm{UL}}_{\mathrm{min}}, \\
& |\theta_n| = 1,\ \forall n = 1,\ldots,N,
\end{aligned}
\end{equation}
where $\boldsymbol{\Phi}=\mathrm{diag}(\theta_1,\ldots,\theta_N)$ and $|\theta_n|=1$.

To solve ~\eqref{eq:sub_phi_original}, we use a successive convex approximation (SCA) method, combining with a trust-region approach, to ensure the approximation accuracy, by parameterizing $\theta_n=e^{j\vartheta_n}$ with $\vartheta_n\in[0,2\pi]$ and letting $\boldsymbol{\vartheta}=[\vartheta_1,\ldots,\vartheta_N]^T$. At iteration $t$, we construct convex lower bounds for the SINR constraints via first--order Taylor expansions around $\boldsymbol{\vartheta}^{(t)}$, denoted by $\Gamma_i^{(t)}(\boldsymbol{\vartheta})$ for $i\in\{\mathrm{DL},\mathrm{UL}\}$. In addition, we linearize the sum rate objective to ensure stable convergence. The resulting convex surrogate problem is
\begin{subequations}\label{eq:phi_sca}
\begin{align}
\max_{\boldsymbol{\vartheta}} \quad &
\left(\nabla_{\boldsymbol{\vartheta}} R_{\mathrm{DL}}(\boldsymbol{\vartheta}^{(t)}) 
+ \nabla_{\boldsymbol{\vartheta}} R_{\mathrm{UL}}(\boldsymbol{\vartheta}^{(t)})\right)^{T}
(\boldsymbol{\vartheta}-\boldsymbol{\vartheta}^{(t)})
\label{eq:phi_sca_a}\\
\text{s.t.}\quad &
\Gamma_i^{(t)}(\boldsymbol{\vartheta}) \ge \gamma_{\mathrm{min},i},
\quad i\in\{\mathrm{DL},\mathrm{UL}\}
\label{eq:phi_sca_b}\\
& 0 \le \vartheta_n \le 2\pi,\quad \forall n=1,\ldots,N
\label{eq:phi_sca_c}\\
& \left\|\boldsymbol{\vartheta}-\boldsymbol{\vartheta}^{(t)}\right\|_2
\le \Delta^{(t)}
\label{eq:phi_sca_d}
\end{align}
\end{subequations}
where \eqref{eq:phi_sca_d} is the trust region constraint with radius $\Delta^{(t)}$. In trust-region-based SCA, this constraint restricts the optimization variables' update to a neighborhood of the current iterate, thereby ensuring the validity of the first-order Taylor approximations and preventing overly aggressive steps that could violate the local convexity assumptions. The trust region radius $\Delta^{(t)}$ is adjusted based on the agreement between the surrogate model and the actual objective. If the approximation is accurate, the radius is increased to allow larger steps. Otherwise, it is reduced to make more conservative updates \cite{trust1}. This method ensures that the surrogate problem provides a reliable local approximation of the original non-convex problem and guarantees stable convergence to a stationary point \cite{trust2, trust3}. After solving problem~\eqref{eq:phi_sca}, we update $\boldsymbol{\vartheta}$ and reconstruct $\boldsymbol{\Phi}$ as $\boldsymbol{\Phi}=\mathrm{diag}(e^{j\vartheta_1},\ldots,e^{j\vartheta_N})$. The detailed explanations for the gradient calculation are discussed in Appendix~\ref{app:phi_grad}.

\subsection{BS transmit antenna locations ($\mathbf{T}_t$)}
In this subsection, we optimize the BS transmit antenna locations $\mathbf{T}_t=\{\mathbf{t}_m\}_{m=1}^{M_t}$ with all other variables fixed. The antenna locations affect the field--response vectors and thus the effective DL and SI channels. The resulting subproblem maximizes the sum rate subject to QoS and geometric constraints and is given by
\begin{subequations}\label{eq:sub_Tt}
\begin{align}
\max_{\mathbf{T}_t} \quad 
& \log_2\!\left(1+\gamma_{\mathrm{DL}}\right)
+ \log_2\!\left(1+\gamma_{\mathrm{UL}}\right)
\label{eq:sub_Tt_obj} \\[2mm]
\text{s.t.}\quad
& \gamma_{\mathrm{DL}} \ge \gamma_{\mathrm{DL}}^{\min}, 
\label{eq:sub_Tt_DLQoS} \\
& \gamma_{\mathrm{UL}} \ge \gamma_{\mathrm{UL}}^{\min}, 
\label{eq:sub_Tt_ULQoS} \\
& \mathbf{t}_m \in \mathcal{T}_m,\quad \forall m \in \{1,\ldots,M_t\}, 
\label{eq:sub_Tt_box} \\
& \|\mathbf{t}_m - \mathbf{t}_{m'}\|_2 \ge d_0, \quad \forall m \neq m'. 
\label{eq:sub_Tt_sep}
\end{align}
\end{subequations}
This subproblem is non--convex due to nonlinear dependence of $\gamma_{\mathrm{DL}}$ and $\gamma_{\mathrm{UL}}$ on $\mathbf{T}_t$. In our implementation, we handle it using a local gradient-based approach using Appendix~\ref{app:pos_grads}, update over the feasible region defined by \eqref{eq:sub_Tt_box}--\eqref{eq:sub_Tt_sep}.

Since $\mathbf{T}_t $ is defined as $ [x_1, \ldots, x_{M_t}; y_1, \ldots, y_{M_t}] \in \mathbb{R}^{2 \times M_t}$, where the first row contains the $x$-coordinates and the second row contains the $y$-coordinates of the $M_t$ transmit antennas, the subproblem is reformulated as
\small
\begin{subequations}\label{eq:Tt_sca}
\begin{align}
\max_{\mathbf{T}_t} \quad &
\mathrm{vec}\left(\nabla_{\mathbf{T}_t} R_{\mathrm{DL}}(\mathbf{T}_t^{(t)}) 
+ \nabla_{\mathbf{T}_t} R_{\mathrm{UL}}(\mathbf{T}_t^{(t)})\right)^{T}
\mathrm{vec}(\mathbf{T}_t-\mathbf{T}_t^{(t)})
\label{eq:Tt_sca_a}\\
\text{s.t.}\quad &
\Gamma_i^{(t)}(\mathbf{T}_t) \ge \gamma_{\mathrm{min},i},
\quad i\in\{\mathrm{DL},\mathrm{UL}\},
\label{eq:Tt_sca_b}\\
& \mathbf{t}_m \in \mathcal{T}_m,\quad \forall m \in \{1,\ldots,M_t\}, 
\label{eq:sub_Tt_box2} \\
& \|\mathbf{t}_m - \mathbf{t}_{m'}\|_2 \ge d_0, \quad \forall m \neq m', 
\label{eq:sub_Tt_sep2}\\
& \left\|\mathrm{vec}(\mathbf{T}_t-\mathbf{T}_t^{(t)})\right\|_2
\le \Delta^{(t)},
\label{eq:Tt_trust}
\end{align}
\end{subequations}
where \eqref{eq:sub_Tt_box2} indicates the feasible region for each antenna, \eqref{eq:sub_Tt_sep2} is the minimum separation distance, and \eqref{eq:Tt_trust} is the trust region constraint.
Also, $\Delta_{t}^{(t)}$ is the trust region radius.
\subsection{BS receive antenna locations ($\mathbf{T}_r$)}
We optimize now the BS receive antenna locations $\mathbf{T}_r=\{\mathbf{u}_m\}_{m=1}^{M_r}$ with all other variables fixed. The receive antenna positions affect the uplink effective channel as well as the residual SI coupling. The corresponding subproblem is
\begin{subequations}\label{eq:sub_Tr}
\begin{align}
\max_{\mathbf{T}_r} \quad 
& \log_2\!\left(1+\gamma_{\mathrm{UL}}\right)
\label{eq:sub_Tr_obj} \\[2mm]
\text{s.t.}\quad
& \gamma_{\mathrm{UL}} \ge \gamma_{\mathrm{UL}}^{\min}, 
\label{eq:sub_Tr_ULQoS} \\
& \mathbf{u}_m \in \mathcal{U}_m,\quad \forall m \in \{1,\ldots,M_r\}, 
\label{eq:sub_Tr_box} \\
& \|\mathbf{u}_m - \mathbf{u}_{m'}\|_2 \ge d_0, \quad \forall m \neq m'. 
\label{eq:sub_Tr_sep}
\end{align}
\end{subequations}
Similar to the $\mathbf{T}_t$ subproblem, this problem is non--convex and is solved via a local optimization based on Appendix~\ref{app:pos_grads} within the feasible set.
\subsection{ME--RIS element locations ($\mathbf{R}$)}
In this subsection, we optimize the ME--RIS element locations $\mathbf{R}=\{\mathbf{r}_n\}_{n=1}^{N}$ with all other variables fixed. The element positions determine the field-response matrices in the BS--RIS, RIS--DL, and RIS--UL links and hence govern the composite channels in both downlink and uplink SINRs. The location optimization subproblem is
\begin{subequations}\label{eq:sub_R}
\begin{align}
\max_{\mathbf{R}} \quad 
& \log_2\!\left(1+\gamma_{\mathrm{DL}}\right)
+ \log_2\!\left(1+\gamma_{\mathrm{UL}}\right)
\label{eq:sub_R_obj} \\[2mm]
\text{s.t.}\quad
& \gamma_{\mathrm{DL}} \ge \gamma_{\mathrm{DL}}^{\min}, 
\label{eq:sub_R_DLQoS} \\
& \gamma_{\mathrm{UL}} \ge \gamma_{\mathrm{UL}}^{\min}, 
\label{eq:sub_R_ULQoS} \\
& \mathbf{r}_n \in \mathcal{R}_n,\quad \forall n \in \{1,\ldots,N\}, 
\label{eq:sub_R_box} \\
& \|\mathbf{r}_n - \mathbf{r}_{n'}\|_2 \ge d_0, \quad \forall n \neq n'.
\label{eq:sub_R_sep}
\end{align}
\end{subequations}
which is again non--convex due to the nonlinear dependence of the field-response channels on $\mathbf{R}$. We solve this subproblem using a local position update method based on Appendix~\ref{app:pos_grads} that guarantees feasibility with respect to \eqref{eq:sub_R_box}--\eqref{eq:sub_R_sep}.
Therefore, the final form of the subproblem solvable by optimization toolboxes such as CVX can be shown by,
\small
\begin{subequations}\label{eq:R_sca}
\begin{align}
\max_{\mathbf{R}} \quad &
\mathrm{vec}\left(\nabla_{\mathbf{R}} R_{\mathrm{DL}}(\mathbf{R}^{(t)}) 
+ \nabla_{\mathbf{R}} R_{\mathrm{UL}}(\mathbf{R}^{(t)})\right)^{T}
\mathrm{vec}(\mathbf{R}-\mathbf{R}^{(t)})
\label{eq:R_sca_a}\\
\text{s.t.}\quad &
\Gamma_i^{(t)}(\mathbf{R}) \ge \gamma_{\mathrm{min},i},
\quad i\in\{\mathrm{DL},\mathrm{UL}\},
\label{eq:R_sca_b}\\
& \mathbf{r}_n \in \mathcal{R}_n,\quad \forall n \in \{1,\ldots,N\}, 
\label{eq:sub_R_box2} \\
& \|\mathbf{r}_n - \mathbf{r}_{n'}\|_2 \ge d_0, \quad \forall n \neq n', 
\label{eq:sub_R_sep2}\\
& \left\|\mathrm{vec}(\mathbf{R}-\mathbf{R}^{(t)})\right\|_2
\le \Delta^{(t)},
\label{eq:R_trust}
\end{align}
\end{subequations}

Finally, the proposed AO algorithm is applied, solving the subproblems iteratively, as summarized in Algorithm~\ref{alg:ao}.
\begin{algorithm}[!t]
\caption{Alternating Optimization (AO) Algorithm}
\label{alg:ao}
\begin{algorithmic}[1]
\State \textbf{Initialize:} set $n=0$, choose a feasible initial point
$\big(\boldsymbol{\omega}^{(0)},\boldsymbol{v}^{(0)},\boldsymbol{\Phi}^{(0)},\mathbf{T}_t^{(0)},\mathbf{T}_r^{(0)},\mathbf{R}^{(0)}\big)$,
tolerance $\epsilon$, and maximum iterations $n_{\max}$.
\State Compute $R_{\mathrm{sum}}^{(0)}$.

\Repeat
    \State Update $\boldsymbol{\omega}^{(n+1)}$ by solving \eqref{eq:SDP}
    for fixed $\{\boldsymbol{v}^{(n)}, p^{(n)}, \boldsymbol{\Phi}^{(n)}, \mathbf{T}_t^{(n)}, \mathbf{T}_r^{(n)}, \mathbf{R}^{(n)}\}$
    (and the auxiliary variables $\boldsymbol{\zeta},\boldsymbol{\beta}$ if applicable).

    \State Update $\boldsymbol{v}^{(n+1)}$ by solving \eqref{eq:sub_v_original}
    for fixed $\{\boldsymbol{\omega}^{(n+1)}, p^{(n)}, \boldsymbol{\Phi}^{(n)}, \mathbf{T}_t^{(n)}, \mathbf{T}_r^{(n)}, \mathbf{R}^{(n)}\}$.

    \State Update $p^{(n+1)}$ by solving \eqref{eq:sub_p_original}
    for fixed $\{\boldsymbol{\omega}^{(n+1)}, \boldsymbol{v}^{(n+1)}, \boldsymbol{\Phi}^{(n)}, \mathbf{T}_t^{(n)}, \mathbf{T}_r^{(n)}, \mathbf{R}^{(n)}\}$.

    \State Update $\boldsymbol{\Phi}^{(n+1)}$ by solving \eqref{eq:phi_sca}
    for fixed $\{\boldsymbol{\omega}^{(n+1)}, \boldsymbol{v}^{(n+1)}, p^{(n+1)}, \mathbf{T}_t^{(n)}, \mathbf{T}_r^{(n)}, \mathbf{R}^{(n)}\}$.

    \State Update $\mathbf{T}_t^{(n+1)}$ by solving \eqref{eq:sub_Tt}
    for fixed $\{\boldsymbol{\omega}^{(n+1)}, \boldsymbol{v}^{(n+1)}, p^{(n+1)}, \boldsymbol{\Phi}^{(n+1)}, \mathbf{T}_r^{(n)}, \mathbf{R}^{(n)}\}$.

    \State Update $\mathbf{T}_r^{(n+1)}$ by solving \eqref{eq:sub_Tr}
    for fixed $\{\boldsymbol{\omega}^{(n+1)}, \boldsymbol{v}^{(n+1)}, p^{(n+1)}, \boldsymbol{\Phi}^{(n+1)}, \mathbf{T}_t^{(n+1)}, \mathbf{R}^{(n)}\}$.

    \State Update $\mathbf{R}^{(n+1)}$ by solving \eqref{eq:sub_R}
    for fixed $\{\boldsymbol{\omega}^{(n+1)}, \boldsymbol{v}^{(n+1)}, p^{(n+1)}, \boldsymbol{\Phi}^{(n+1)}, \mathbf{T}_t^{(n+1)}, \mathbf{T}_r^{(n+1)}\}$.

    \State Compute $R_{\mathrm{sum}}^{(n+1)}$ and set $n \leftarrow n+1$.
\Until{$R_{\mathrm{sum}}^{(n)} - R_{\mathrm{sum}}^{(n-1)}\le \epsilon$ \textbf{or} $n \ge n_{\max}$}

\State \textbf{Output:}
$\mathcal{X}^\star \leftarrow \{p^{(n)},\boldsymbol{\omega}^{(n)},\boldsymbol{v}^{(n)},\boldsymbol{\Phi}^{(n)},\mathbf{T}_t^{(n)},\mathbf{T}_r^{(n)},\mathbf{R}^{(n)}\}$.
\end{algorithmic}
\end{algorithm}

\subsection{Computational Complexity Analysis}

We analyze the computational complexity of the proposed alternating optimization algorithm. Let $I_{\text{AO}}$ denote the number of AO iterations required for convergence.

For the downlink beamforming ($\boldsymbol{\omega}$) optimization, we employ the SDR-SROCR method by solving a semidefinite programming (SDP) problem at each SROCR iteration. According to \cite{cvx, semidefinite}, the worst-case computational complexity of solving a generic SDP by the CVX toolbox is given by $\mathcal{O}(\max\{m, n\}^4 n^{1/2} \log(1/\epsilon_1))$, where $m$ is the number of constraints, $n$ is the size of the positive semidefinite matrix, and $\epsilon_1 > 0$ reflects the solution accuracy. For problem (35), we have $m = \mathcal{O}(1)$ and $n = M_t + 1$. Let $I_{\text{SROCR}}$ denote the number of SROCR iterations. Since the product of two positive integers tends to grow faster than a linear increase, the complexity is simplified to $\mathcal{O}(I_{\text{SROCR}} (M_t + 1)^{4.5} \log(1/\epsilon_1))$.

The uplink combiner is obtained via generalized eigenvalue decomposition with complexity $\mathcal{O}(M_r^3)$ \cite{golub}. In addition, the uplink power optimization uses bisection search with $I_{\text{bisect}} = \mathcal{O}(\log_2(P_{u}^{\text{max}}/\epsilon_p))$ iterations, each requiring SINR evaluations with complexity $\mathcal{O}(M_t M_r N)$. Moreover, the phase shift optimization employs SCA with $I_{\text{SCA}}^{\Phi}$ iterations, each solving a convex problem with complexity $\mathcal{O}(N^{3.5})$ \cite{boyd}. The position optimizations for BS transmit antennas, BS receive antennas, and RIS elements have complexities $\mathcal{O}(I_{\text{SCA}}^{T_t} M_t N L)$, $\mathcal{O}(I_{\text{SCA}}^{T_r} M_r N L)$, and $\mathcal{O}(I_{\text{SCA}}^{R} N^2 L)$, respectively, where $L = \max\{L_{\text{BR}}, L_{\text{RB}}, L_{\text{Bd}}, L_{\text{Bu}}, L_{\text{Rd}}, L_{\text{Ru}}\}$.

Therefore, the overall computational complexity per AO iteration is
\begin{equation}
\begin{aligned}
\mathcal{O}\Big(
& I_{\text{SROCR}} (M_t+1)^{4.5}\log(1/\epsilon_1)
+ M_r^3
+ I_{\text{bisect}} M_t M_r N
\\
& + I_{\text{SCA}}^{\Phi} N^{3.5}
+ (I_{\text{SCA}}^{T_t} M_t + I_{\text{SCA}}^{T_r} M_r
+ I_{\text{SCA}}^{R} N) N L
\Big)
\end{aligned}
\end{equation}

where the dominant terms are the SDP beamforming optimization and phase shift optimization when $M_t$ and $N$ are large. The fixed-antenna/element benchmarks have lower complexity by eliminating position optimization terms, though at the cost of reduced performance, as shown in Section V.
\section{Numerical Results}
In this section, we evaluate the performance of the proposed full-duplex MA--BS and ME--RIS assisted system through numerical simulations. In the simulations, we consider an MA-BS at $[5, 0, 15],$ an ME-RIS at the location of $[0, 10, 10]$, and an uplink and a downlink user at $[65, 60, 1.5]$ and $[5, 80, 1.5]$, respectively. We assume that the AoDs and AoAs are independent and identically distributed random variables within the interval of $[0,2\pi]$ for all channels. In addition, the numerical results are averaged over 100 independent and random channel realizations. For comparison, we consider four scenarios: 1) \emph{MA--BS and ME--RIS}, where both the BS antennas and RIS elements are movable; 2) \emph{FA--BS and ME--RIS}, where the BS antennas are fixed and the RIS elements are movable; 3) \emph{MA--BS and FE--RIS}, where the BS antennas are movable and the RIS elements are fixed; and 4) \emph{FA--BS and FE--RIS}, where both the BS antennas and RIS elements are fixed. Unless otherwise stated, all schemes are evaluated under the maximum transmit power of 37 dBm at the BS and 20 dBm at the uplink user. Also, the number of paths for all links is assumed to be $L_{\text{BR}}, L_{\text{RB}}, L_{\text{Bd}}, L_{\text{Bu}}, L_{\text{Rd}}, L_{\text{Ru}}= L = 6$, and other parameter values are mentioned in Table~\ref{tab:sim_params}.

\begin{table}[t]
\centering
\caption{Simulation parameters}
\label{tab:sim_params}
\begin{tabular}{|c|c|c|c|}
\hline
\textbf{Parameter} & \textbf{Value} & \textbf{Parameter} & \textbf{Value} \\
\hline
$N_{t}, \, N_{r}$ & $8, 4$ & $\beta_0$ & $-30~\mathrm{dB}$ \\
\hline
$\alpha_{\mathrm{BR}}, \alpha_{\mathrm{Rd}}, \alpha_{\mathrm{Ru}},$ & $2.1, 2.2, 2.2$ & $\alpha_{\mathrm{Bd}}, \alpha_{\mathrm{Bu}}$ & $3.5, 3.5$ \\
\hline
$\lambda,\, A,\, d_0$ & $0.1,\, 4\lambda,\, \lambda/2$ & $\alpha_{\mathrm{ud}}$ & $ 3.7$ \\
\hline
$R_{th,d},\, R_{th,u}$ & $1~\mathrm{bps/Hz}$ & SI & $-90~\mathrm{dB}$ \\
\hline
$\sigma_d^2,\, \sigma_u^2$ & $-174~\mathrm{dBm}/\mathrm{Hz}$ & $P_{BS}^{\max}$ & $37~\mathrm{dBm}$ \\
\hline
$P_u^{\max}$ & $20~\mathrm{dBm}$ & $\epsilon$ & $0.001$ \\
\hline
\end{tabular}
\end{table}

\begin{figure}[!t]
    \centering
    \includegraphics[width=0.8\linewidth]{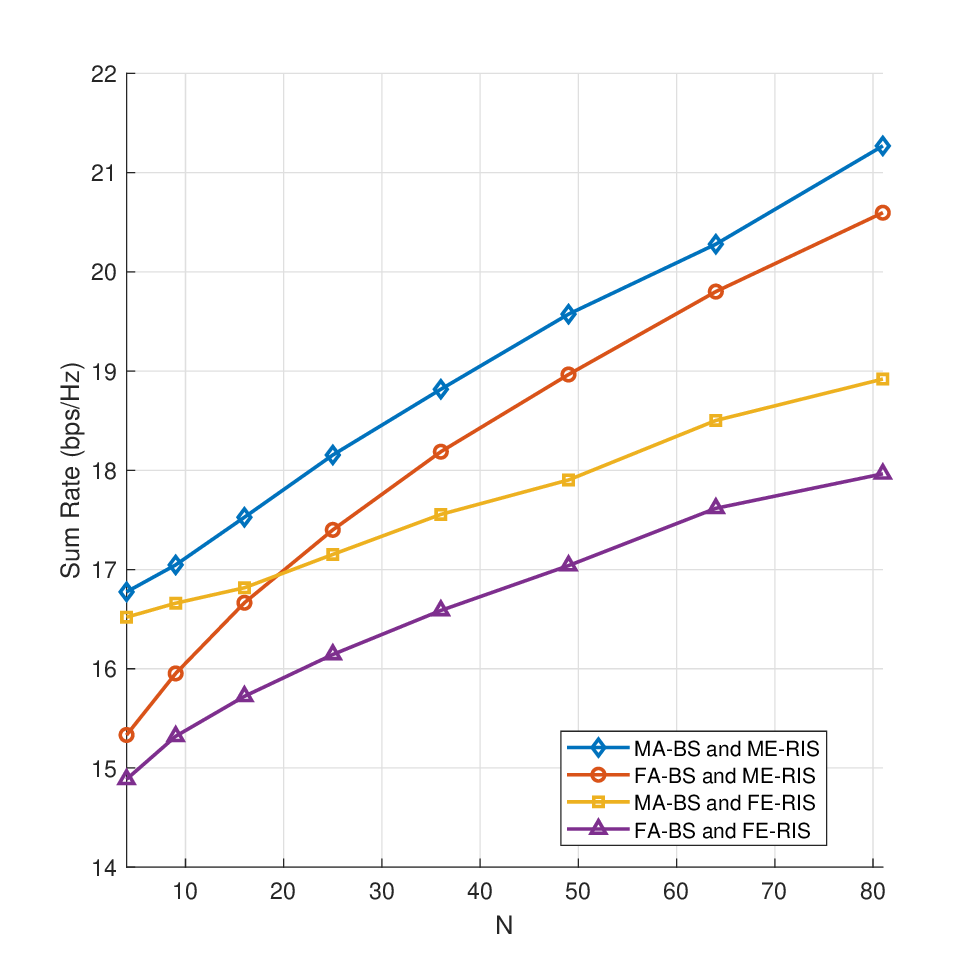}
    \caption{Sum rate versus the number of RIS elements $N$.}
    \label{fig:sumrate_vs_N}
\end{figure}

Fig.~\ref{fig:sumrate_vs_N} illustrates the achievable sum rate as a function of the number of RIS elements $N$. As expected, the sum rate increases monotonically with $N$ for all schemes, because a larger RIS size provides higher passive beamforming gain and more spatial degrees of freedom to improve the cascaded channels.
The proposed \emph{MA--BS and ME--RIS} scheme achieves the highest sum rate and shows an improvement of $18\%$ at $N=81$ compared to the \emph{FA--BS and FE--RIS} scheme. This is because jointly optimizing the BS antenna positions and RIS element locations enables more effective phase alignment over the BS--RIS--user links and allows the additional RIS elements to contribute more constructively as $N$ increases, thereby improving both signal enhancement and interference management.
However, scenarios with fixed antennas and/or fixed RIS elements benefit less from increasing $N$ due to their limited spatial flexibility, which prevents them from fully using the additional RIS elements. Moreover, as $N$ increases, RIS element mobility has a stronger impact than BS antenna mobility, since relocating a larger number of RIS elements directly improves the dominant cascaded BS--RIS--user links, whereas the performance gain from BS antenna movement does not scale as effectively with $N$. Consequently, newly added RIS elements in these schemes provide less effective phase alignment and interference mitigation, which explains the smaller performance improvement compared to the fully movable configuration.

\begin{figure}[!t]
    \centering
    \includegraphics[width=0.8\linewidth]{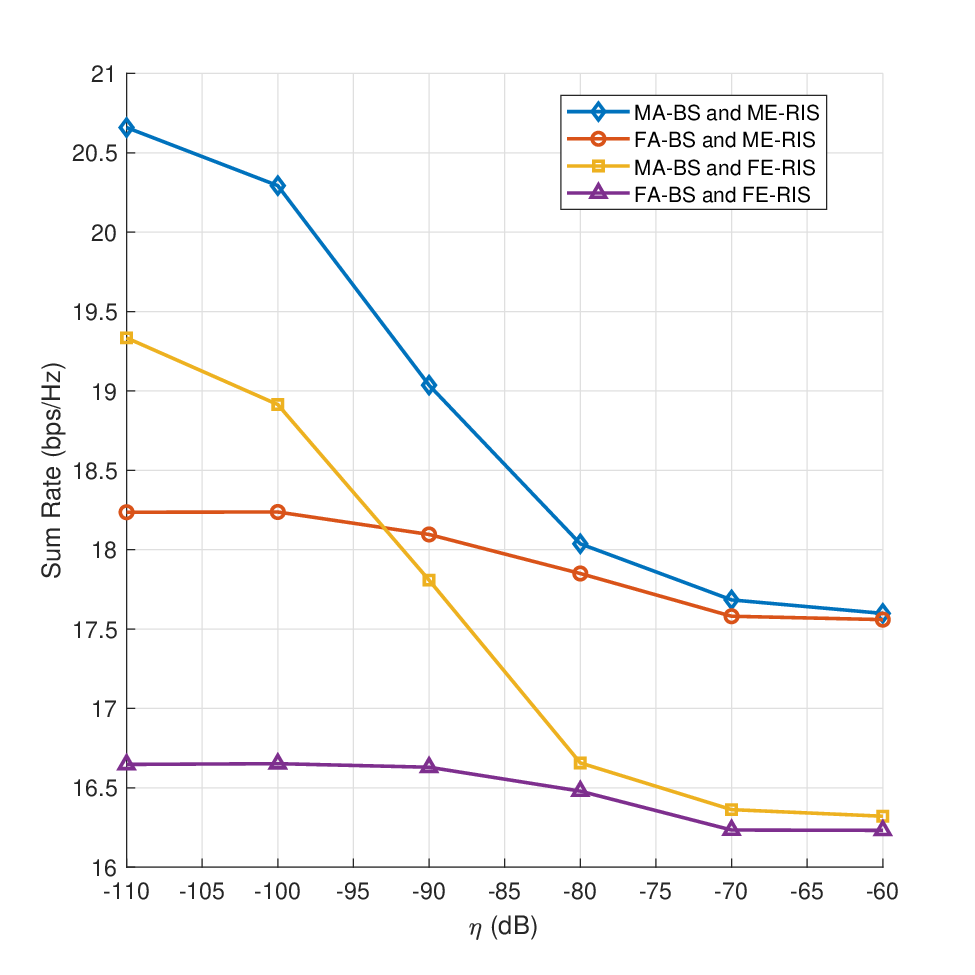}
    \caption{Sum rate versus residual self-interference level $\eta$.}
    \label{fig:sumrate_vs_eta}
\end{figure}

Moreover, Fig.~\ref{fig:sumrate_vs_eta} shows the sum rate versus the SI level $\eta$. As $\eta$ increases, the sum rate decreases for all schemes because higher residual SI worsens the uplink reception at the full-duplex BS.
In the low-SI regime, the proposed \emph{MA--BS and ME--RIS} scheme significantly outperforms the benchmark schemes, demonstrating that it can simultaneously improve the desired links while keeping residual SI under control through spatial reconfiguration.
As $\eta$ increases, the performance gap among schemes gets smaller, indicating that high residual SI becomes the dominant limiting factor, which reduces the achievable gains from mobility and RIS, and the movement of the BS antennas no longer has a beneficial impact. In addition, schemes with movable BS antennas experience a faster performance degradation as $\eta$ increases, since the spatially optimized transmit and reflection paths also amplify the residual SI components. In contrast, schemes with fixed BS antennas show a more gradual performance decrease, because their limited spatial flexibility results in weaker coupling between the desired signal enhancement and SI amplification.

\begin{figure}[!t]
    \centering
    \includegraphics[width=0.8\linewidth]{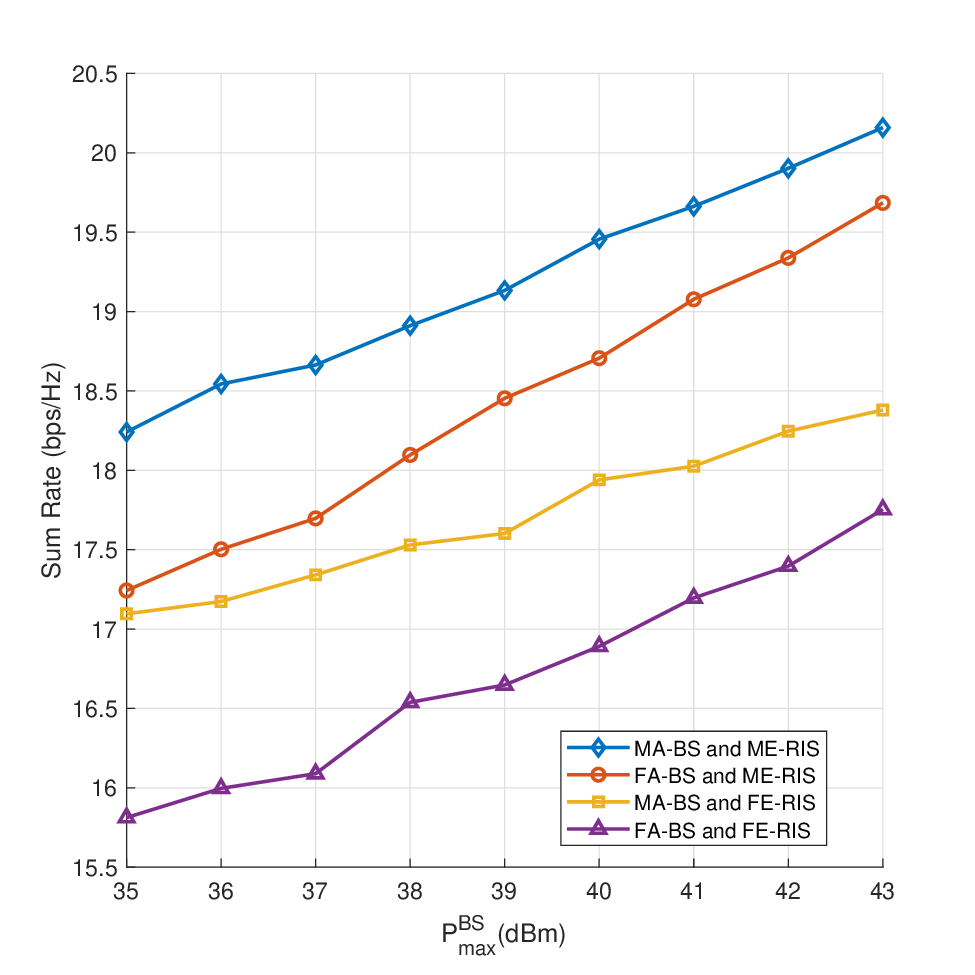}
    \caption{Sum rate versus the maximum transmit power at the BS, $P^{\text{BS}}_{\max}$.}
    \label{fig:sumrate_vs_pmax}
\end{figure}

Furthermore, Fig.~\ref{fig:sumrate_vs_pmax} shows the achievable sum rate vs the maximum transmit power at the BS $P^{\text{BS}}_{\max}$. It can be seen that the sum rate increases with $P^{\text{BS}}_{\max}$ for all schemes, since larger transmit power improves the downlink received signal strength, and by the joint optimization, it can also improve the overall system performance. Additionally, across the entire range of $P^{\text{BS}}_{\max}$, the \emph{MA--BS and ME--RIS} configuration achieves the best performance, which confirms that combining movable BS antennas with movable RIS elements provides the largest spatial degrees of freedom for improving the desired channels and managing interference, and has an approximately $15 \%$ improvement at $P^{\text{BS}}_{\max} = 40 ~\mathrm{dBm}$.

\begin{figure}[!t]
    \centering
    \includegraphics[width=0.8\linewidth]{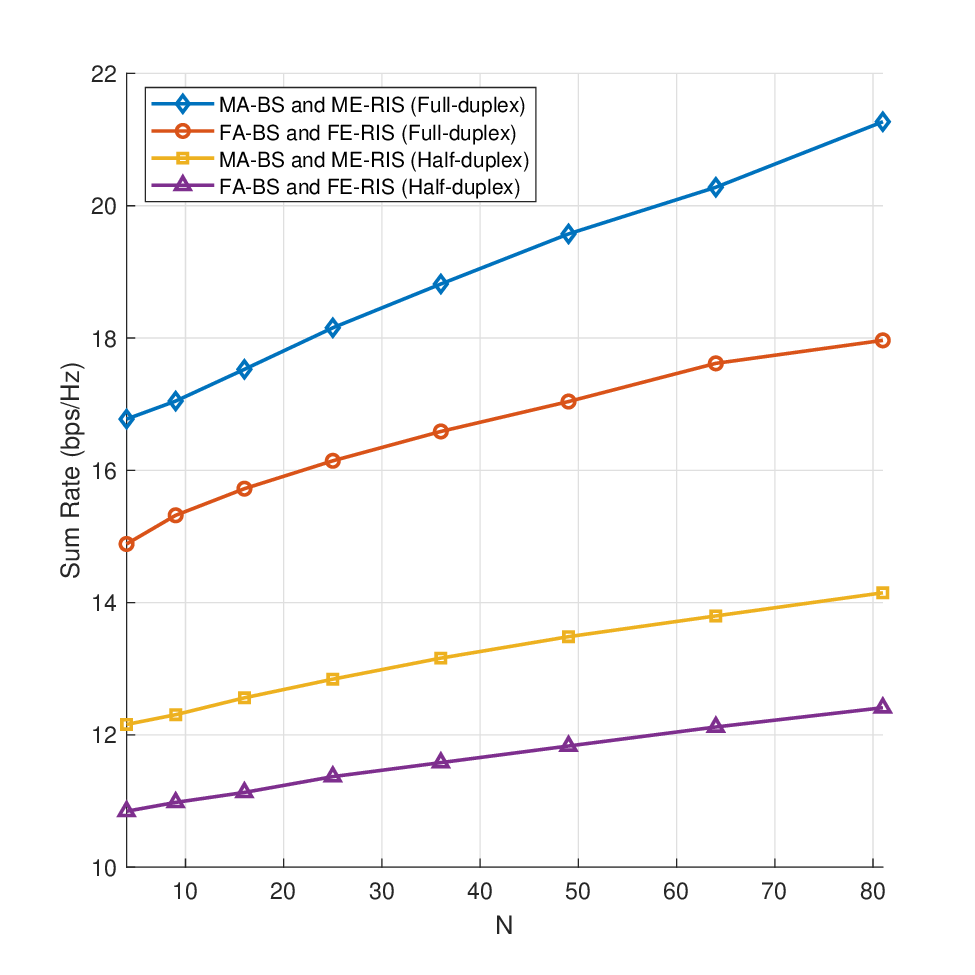}
    \caption{Sum rate for the full--duplex and the half--duplex networks.}
    \label{fig:FDvsHD}
\end{figure}

Fig.~\ref{fig:FDvsHD} compares the sum rate performance of half-duplex and full-duplex systems under \emph{MA--BS and ME--RIS} and \emph{FA--BS and FE--RIS} scenarios. As expected, a full-duplex system achieves higher sum rates than half-duplex transmission, since uplink and downlink signals are transmitted simultaneously rather than in separate time slots. Moreover, the \emph{MA--BS and ME--RIS} setup performs better in both duplexing modes compared to the fixed-position setup due to the additional spatial flexibility provided by movable antennas and RIS elements, which enables more effective enhancement of the desired links and improved interference management. In addition, as the number of RIS elements increases, the performance gains increase as well. This is because a larger RIS provides more spatial degrees of freedom, and the available spatial flexibility becomes more effective in leveraging these degrees of freedom to improve spectral efficiency. As a result, the performance gap between the full-duplex and half-duplex systems, as well as between the movable and fixed configurations, becomes more significant at larger values of $N$.

\begin{figure}[!t]
    \centering
    \includegraphics[width=0.8\linewidth]{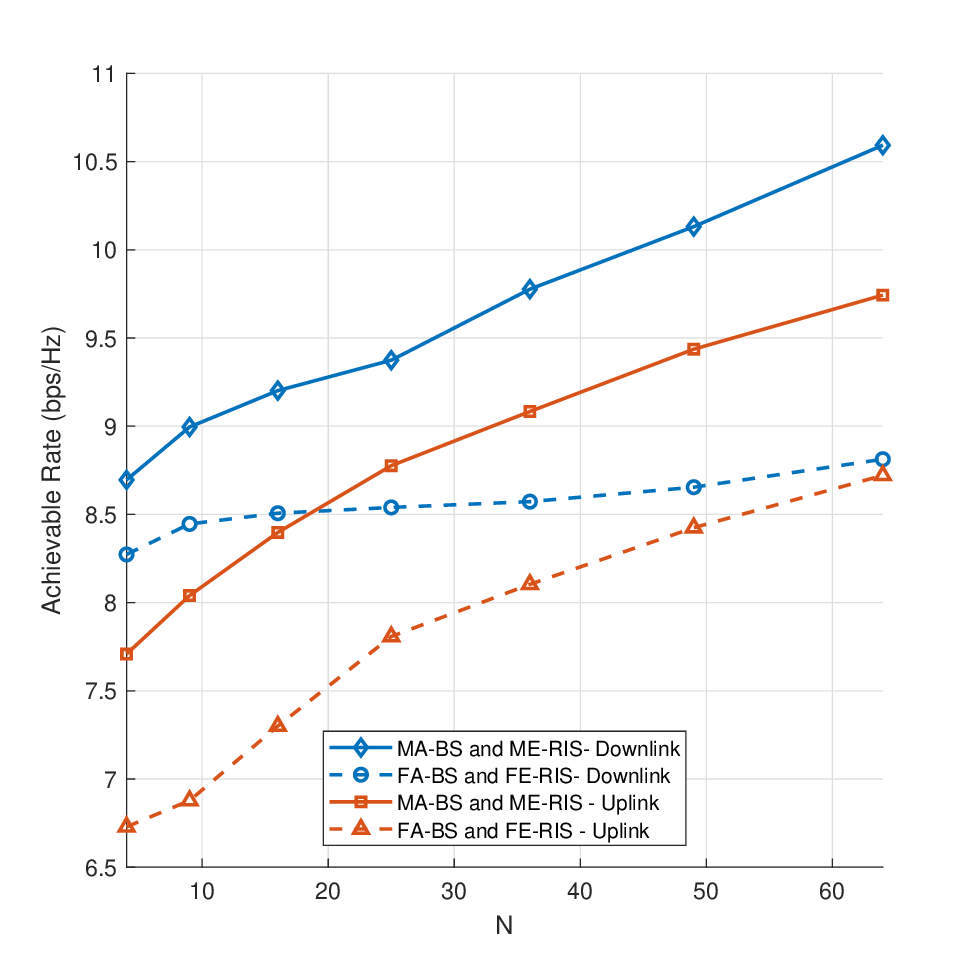}
    \caption{Achievable rate for uplink and downlink users in a full-duplex network.}
    \label{fig:DL_UL}
\end{figure}

Fig.~\ref{fig:DL_UL} shows the achievable rate of the considered full-duplex system as a function of the number of RIS elements $N$ for both the downlink (DL) and uplink (UL). Residual self-interference caused by the BS transmit leakage is included in the uplink SINR, while downlink users are affected by co-channel interference from uplink transmissions.

As $N$ increases, the achievable rate improves in all cases due to the increased passive beamforming gain provided by the RIS. However, the amount of improvement strongly depends on whether the BS antennas and RIS elements are movable or fixed. In the fixed-position case, the downlink rate increases only slightly with $N$, since uplink-to-downlink interference cannot be effectively mitigated using phase-only RIS control when the system geometry is fixed.

In contrast, the proposed architecture with movable BS antennas and movable RIS elements consistently achieves higher rates in both DL and UL. Specifically, at $N=64$, the movable architecture improves the downlink rate by approximately $20\%$ compared to the fixed deployment. The uplink also benefits from mobility, achieving an improvement of about $12\%$ at the same RIS size. These gains highlight the advantage of introducing spatial mobility, which provides additional geometric degrees of freedom to strengthen desired links and better suppress co-channel interference.

Finally, the uplink rates are generally lower than the downlink rates due to residual self-interference at the BS receiver, which creates an interference floor in the uplink SINR. Nevertheless, the movable architecture significantly improves uplink performance by improving the effective uplink channels and interference management, thereby partially reducing the impact of residual self-interference.

\begin{figure}[!t]
    \centering
    \includegraphics[width=0.8\linewidth]{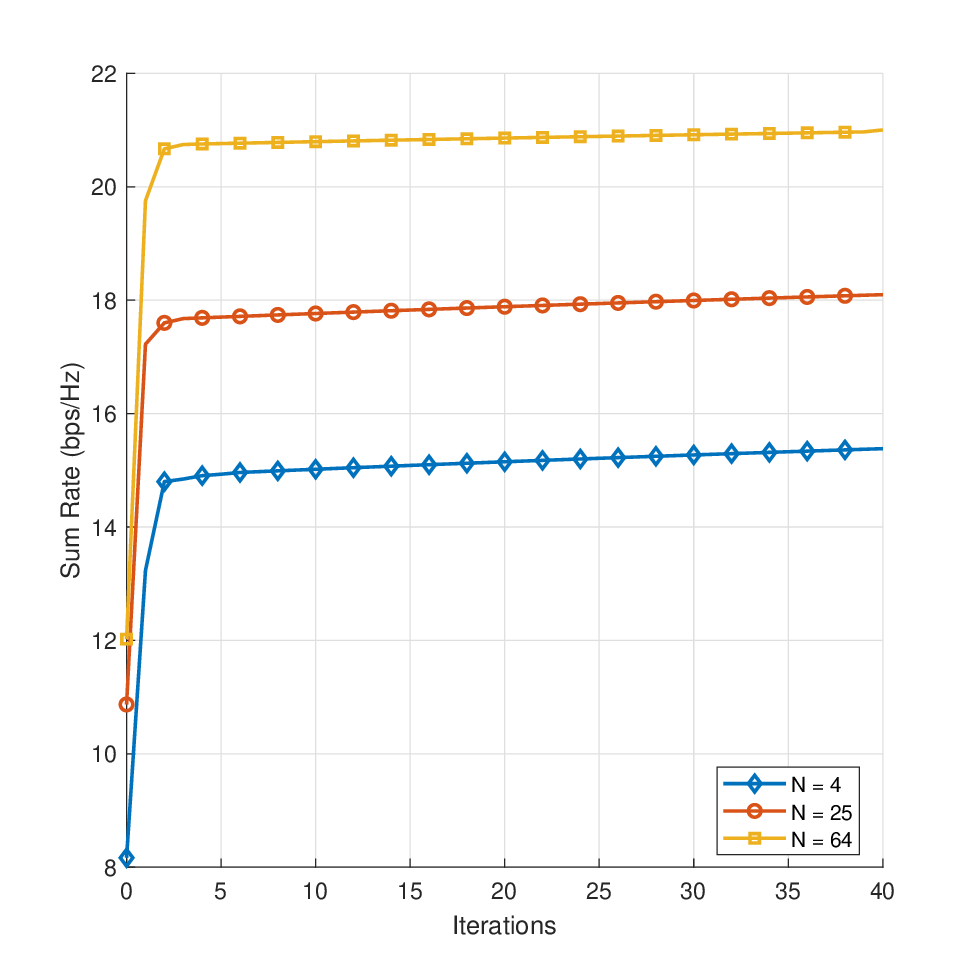}
    \caption{Convergence behavior of the proposed algorithm for different $N$ values}
    \label{fig:convergence}
\end{figure}

Fig.~\ref{fig:convergence} shows the convergence behavior of the proposed algorithm for different $N$ values. It can be seen that the AO algorithm converges after some iterations at AO with six subproblems for all cases.

\section{Conclusion}
In this paper, we investigate the combination of movable antennas at the base station with a movable-element RIS to improve full-duplex communication systems. The key idea is that by letting both the base station antennas and RIS elements move to better positions, we can enhance the intended signals while reducing the negative effects of self-interference and inter-user interference.

We implemented an optimization method that jointly optimizes the beamforming vectors, transmission power, RIS phase shifts, and the physical positions of all movable elements/antennas. Also, the simulation results show that this method outperforms the traditional systems, which have fixed positions for the antennas/elements. The gains are especially significant when the number of RIS elements increases or when self-interference is well-controlled.

It can be concluded that the integration of movable antennas and movable RIS elements offers a promising technique for next-generation wireless networks, making full-duplex systems more practical and efficient.

\appendices

\section{Proof of Lemma~1}
\label{appendix1}

According to the transformed objective function in Eq.~\eqref{LDT1}, it is observed that
$f(\boldsymbol{\omega}, \boldsymbol{\zeta})$ is a concave and differentiable
function with respect to $\boldsymbol{\zeta}$ when $\boldsymbol{\omega}$, is held fixed.
Therefore, the closed-form optimal solution for the auxiliary variables
$\boldsymbol{\zeta}$ can be obtained by setting
$\nabla_{\zeta_{i}} f(\boldsymbol{\zeta}) = 0$, $\forall i \in \{{\mathrm{DL}}, {\mathrm{UL}}\}$.
As a result, the optimal solution is given by
$\zeta^{\star}_{\mathrm{DL}} = \gamma_{\mathrm{DL}}$ and $\zeta^{\star}_{\mathrm{UL}} = \gamma_{\mathrm{UL}}$.

By substituting $\zeta^{\star}_{\mathrm{DL}}$ and $\zeta^{\star}_{\mathrm{UL}}$ into the transformed
objective function $f(\boldsymbol{\omega}, \boldsymbol{\zeta})$ in
Eq.~\eqref{LDT1}, the original sum-rate expression
$\log
_2(1+\gamma_{\mathrm{DL}}) +  \log_2(1+\gamma_{\mathrm{UL}})$ is recovered.
Hence, the equivalence between the transformed objective and the original sum-rate function is established, which completes the proof.

\section{Proof of Lemma~2}
\label{appendix2}

The reformulated function $g(\boldsymbol{\omega}, \boldsymbol{\zeta},
\boldsymbol{\beta})$ introduced in Eq.~\eqref{eq:QT1} is a concave and differentiable
function with respect to the auxiliary variables $\boldsymbol{\beta}$ for fixed
$\boldsymbol{\omega}$, and $\boldsymbol{\zeta}$.
Therefore, the closed-form optimal values of $\beta_{\mathrm{DL}}$, and $\beta_{\mathrm{UL}}$,
can be obtained by solving
$\nabla_{\beta_{i}} g(\boldsymbol{\beta}) = 0, ~ i \in \{{\mathrm{DL}}, {\mathrm{UL}}\} $.

The optimal solution is given by
\begin{align}
\beta_{\mathrm{DL}}^{*}
&=
\frac{\sqrt{1+\zeta_{\mathrm{DL}}}\,
\big(\boldsymbol{h}_{\mathrm{d}}
+
\tilde{\boldsymbol{h}}\boldsymbol{\Phi}\boldsymbol{H}\big)\boldsymbol{\omega}
}{{
\big|\big(\boldsymbol{h}_{\mathrm{d}}
+\tilde{\boldsymbol{h}}\boldsymbol{\Phi}\boldsymbol{H}\big)\boldsymbol{\omega}\big|^2
} + {
p\,\big| I+\tilde{\boldsymbol{h}}\boldsymbol{\Phi}\boldsymbol{g} \big|^2
+\sigma_d^2
}
},
\\[2mm]
\beta_{\mathrm{UL}}^{*}
&=
\frac{\sqrt{p(1+\zeta_{\mathrm{UL}})}\,\,
\boldsymbol{v}^{H}
\big(\boldsymbol{h}_{\mathrm{u}}
+
\boldsymbol{G}\boldsymbol{\Phi}\boldsymbol{g}\big)
}{
p\,\big|\boldsymbol{v}^{H}
\big(\boldsymbol{h}_{\mathrm{u}}
+
\boldsymbol{G}\boldsymbol{\Phi}\boldsymbol{g}\big)\big|^{2}
+
\eta\,
\big|\boldsymbol{v}^{H}
\big(\boldsymbol{F}
+
\boldsymbol{G}\boldsymbol{\Phi}\boldsymbol{H}\big)\boldsymbol{\omega}\big|^{2}
+
\sigma_u^{2}\|\boldsymbol{v}\|_{2}^{2}
}.
\end{align}

Substituting $\beta_{\mathrm{DL}}^{*}$ and $\beta_{\mathrm{DL}}^{*}$ into
$g(\boldsymbol{\omega}, \boldsymbol{\zeta}, \boldsymbol{\beta})$ accurately
recovers the multiple-ratio fractional programming (MRFP) term
$\sum_{i \in \{{\mathrm{DL}}, {\mathrm{UL}}\}}(1+\zeta_{i}) f_{i}(\boldsymbol{\omega})$ defined in Eq.~\eqref{LDT1}. Hence, the equivalence is established, and the proof is complete.

\section{ Position-Gradient Formulas}
\label{app:pos_grads}

\subsection{FRV/FRM and Cascaded-Channel Derivatives}
Let \(\mathbf{p}=[x,y]^T\) denote a movable antenna/element position and define the generic FRV
\begin{equation}
\mathbf{a}(\mathbf{p})=\big[e^{j\frac{2\pi}{\lambda}\rho_1(\mathbf{p})},\ldots,e^{j\frac{2\pi}{\lambda}\rho_L(\mathbf{p})}\big]^T,
\quad
\rho_\ell(\mathbf{p})=x\kappa_{\ell,x}+y\kappa_{\ell,y},
\end{equation}
where \(\kappa_{\ell,x}\triangleq \cos(\theta_\ell)\sin(\phi_\ell)\) and \(\kappa_{\ell,y}\triangleq \sin(\theta_\ell)\).
Let \(\boldsymbol{\kappa}_x=[\kappa_{1,x},\ldots,\kappa_{L,x}]^T\) and \(\boldsymbol{\kappa}_y=[\kappa_{1,y},\ldots,\kappa_{L,y}]^T\).
Then
\begin{equation}
\frac{\partial \mathbf{a}}{\partial x}=j\frac{2\pi}{\lambda}\operatorname{diag}(\boldsymbol{\kappa}_x)\mathbf{a},
\qquad
\frac{\partial \mathbf{a}}{\partial y}=j\frac{2\pi}{\lambda}\operatorname{diag}(\boldsymbol{\kappa}_y)\mathbf{a}.
\label{eq:app_frv_grad}
\end{equation}
For an FRM \(\mathbf{E}=[\mathbf{a}(\mathbf{p}_1), \ldots, \mathbf{a}(\mathbf{p}_M)]\), the derivative w.r.t. \((x_m,y_m)\) affects only column \(m\):
\(\frac{\partial \mathbf{E}}{\partial x_m}=[\mathbf{0},\ldots,\frac{\partial \mathbf{a}(\mathbf{p}_m)}{\partial x},\ldots,\mathbf{0}]\),
and similarly for \(\partial \mathbf{E}/\partial y_m\). The same applies to receive-side FRMs, \(\mathbf{F}\), w.r.t. \((x_n,y_n)\).
For a generic FR channel \(\mathbf{H}=\mathbf{F}^H\boldsymbol{\Sigma}\mathbf{E}\),
\begin{equation}
\frac{\partial \mathbf{H}}{\partial q}=
\begin{cases}
\mathbf{F}^H\boldsymbol{\Sigma}\,\frac{\partial \mathbf{E}}{\partial q}, & q\in\{x_m,y_m\},\\[1mm]
\left(\frac{\partial \mathbf{F}}{\partial q}\right)^H\boldsymbol{\Sigma}\mathbf{E}, & q\in\{x_n,y_n\},
\end{cases}
\label{eq:app_H_grad}
\end{equation}
where \(q\) is any scalar position coordinate of antenna/element at the transmitter or the receiver side.
\subsection{SINR and Rate Gradients}
Define the general SINR formula as
\begin{equation}
\gamma=\frac{|s|^2}{|I|^2 + n}.
\label{eq:app_dl_defs}
\end{equation}
For any scalar position variable \(q\in\{x,y\}\),
\begin{equation}
\frac{\partial \gamma}{\partial q}
=
\frac{2|I|^2}{(|I|^2+n)^2}\Re\!\left\{s^*\frac{\partial s}{\partial q}\right\}
-\frac{2|s|^2}{(|I|^2+n)^2}\,\Re\!\left\{I^*\frac{\partial I}{\partial q}\right\},
\label{eq:app_grad_gamma_dl}
\end{equation}
assuming $s = (\boldsymbol{h}_1 + \boldsymbol{h}_2 \boldsymbol{\Phi} \boldsymbol{H}_1) \boldsymbol{v}_1$, and $I = (\boldsymbol{h}_3 + \boldsymbol{h}_4 \boldsymbol{\Phi} \boldsymbol{H}_2)\boldsymbol{v}_2$, the derivatives can be done as follows,
\begin{align}
\frac{\partial s}{\partial q}
=
\left(\frac{\partial \mathbf{h}_1}{\partial q} + \frac{\partial}{\partial q}\!\left({\mathbf{h}_2}\boldsymbol{\Phi}\mathbf{H}_1\right)\right)\boldsymbol{v}_1,
\\
\frac{\partial I}{\partial q}
=
\left(\frac{\partial \mathbf{h}_3}{\partial q} + \frac{\partial}{\partial q}\!\left({\mathbf{h}_4}\boldsymbol{\Phi}\mathbf{H}_2\right)\right)\boldsymbol{v}_2.
\label{eq:app_grad_s_I}
\end{align}
where
\begin{align}
\frac{\partial}{\partial q}\!\left({\mathbf{h}}\boldsymbol{\Phi}\mathbf{H}\right)
&=
\frac{\partial {\mathbf{h}}}{\partial q}\boldsymbol{\Phi}\mathbf{H}
+ {\mathbf{h}}\boldsymbol{\Phi}\frac{\partial \mathbf{H}}{\partial q},
\label{eq:app_grad_prod1}
\end{align}
where \(\partial \mathbf{H}/\partial q\) follows \eqref{eq:app_H_grad}, and \(\partial {\mathbf{h}}/\partial q\) follows \eqref{eq:app_frv_grad} by applying the same FRV/FRM construction for the corresponding links.
Finally, for \(R =\log_2(1+\gamma)\),
\begin{equation}
\frac{\partial R}{\partial q}
=
\frac{\frac{\partial \gamma}{\partial q}}{(\ln 2)(1+\gamma)}.
\label{eq:app_grad_Rdl}
\end{equation}

\section{ Phase shift-Gradient Formulas}
\label{app:phi_grad}

Define the general SINR formula as
\begin{equation}
\gamma=\frac{|s|^2}{|I|^2 + n}.
\label{eq:app_dl_defs2}
\end{equation}
By reformulating the phase shift variable $\boldsymbol{\Phi}$ as a row vector of $\boldsymbol{\theta} = \mathrm{diag(\boldsymbol{\Phi})}$, we will have the derivatives of the general SINR for the gradient calculation as,
\begin{equation}
\frac{\partial \gamma}{\partial \boldsymbol{\theta}}
=
\frac{2|I|^2}{(|I|^2+n)^2}\Re\!\left\{s^*\frac{\partial s}{\partial \boldsymbol{\theta}}\right\}
-\frac{2|s|^2}{(|I|^2+n)^2}\,\Re\!\left\{I^*\frac{\partial I}{\partial \boldsymbol{\theta}}\right\},
\label{eq:app_grad_gamma_dl2}
\end{equation}
assuming $s = (\boldsymbol{h}_1 + \boldsymbol{h}_2 \boldsymbol{\Phi} \boldsymbol{H}_1) \boldsymbol{v}_1$, and $I = (\boldsymbol{h}_3 + \boldsymbol{h}_4 \boldsymbol{\Phi} \boldsymbol{H}_2)\boldsymbol{v}_2$, after applying the reformulations we will have $s = (\boldsymbol{h}_1 + \boldsymbol{\theta}\,\mathrm{diag}(\boldsymbol{h}_2) \boldsymbol{H}_1) \boldsymbol{v}_1$, and $I = (\boldsymbol{h}_3 + \boldsymbol{\theta}\, \mathrm{diag}(\boldsymbol{h}_4) \boldsymbol{H}_2)\boldsymbol{v}_2$. Therefore, the derivatives can be done as follows,
\begin{align}
\frac{\partial s}{\partial \boldsymbol{\theta}}
=
\left(\mathrm{diag}(\boldsymbol{h}_2) \boldsymbol{H}_1\right) \boldsymbol{v}_1,
\\
\frac{\partial I}{\partial \boldsymbol{\theta}}
=
\left(\mathrm{diag}(\boldsymbol{h}_4) \boldsymbol{H}_2\right)\boldsymbol{v}_2.
\label{eq:app_grad_s_I2}
\end{align}
Finally, we apply the chain rule to calculate the gradients w.r.t. $\boldsymbol{\vartheta}$, as we defined $\theta_n=e^{j\vartheta_n}$.

\bibliographystyle{IEEEtran}
\bibliography{references.bib}


\newpage

 




\vfill

\end{document}